\documentclass{sig-alternate}
\usepackage{graphicx}
\usepackage{subfigure}
\usepackage{algorithmic}
\usepackage{amssymb}
\usepackage{url}
\usepackage{algorithm}
\newcommand{\eg}{{\it e.g.}\xspace}
\newcommand{\ie}{{\it i.e.}\xspace}

\newcommand{\vs}{{\it vs.}\xspace}
\usepackage{color}
\usepackage{ifthen}
\newboolean{TECHREPORT}
\setboolean{TECHREPORT}{false}
\usepackage{cite}
\usepackage{amssymb,amsmath}
\usepackage{newclude}
\usepackage{epstopdf}
\usepackage{comment}
\usepackage{multirow}
\usepackage{xspace}
\usepackage{pdfpages}

\usepackage[T1]{fontenc}
\usepackage[latin2]{inputenc}
\usepackage{balance}

\definecolor{heraldBlue}{rgb}{0.0,0.0,0.8}
\definecolor{heraldRed}{rgb}{0.8,0.0,0.0}
\definecolor{heraldGray}{rgb}{0.4,0.4,0.4}
\definecolor{heraldBlack}{rgb}{0.0,0.0,0.0} 
\definecolor{heraldGreen}{rgb}{0.0,0.4,0.0} 

\newcommand{\hwmatching}{{\tt End-Points Matching}\xspace}
\newcommand{\enroutematching}{{\tt En-Route Matching}\xspace}

\newcommand{\hwonly}{{\tt End-Points RS}\xspace}
\newcommand{\fullroute}{{\tt En-Route RS}\xspace}

\begin{document}

\title{Assessing the Potential of Ride-Sharing Using Mobile and Social Data}
\subtitle{A Tale of Four Cities}

\numberofauthors{1}
\author{  \alignauthor  Blerim Cici$^{\dagger\star}$, Athina Markopoulou$^\dagger$,  Enrique Fr\'{\i}as-Mart\'{\i}nez$^\star$,  Nikolaos Laoutaris$^\star$\\
        \affaddr{UC Irvine$^\dagger$, Telefonica Research(Spain)$^\star$}\\
	\email{ \{bcici, athina\}@uci.edu, \{efm, nikos\}@tid.es}
}

\maketitle

\begin{abstract}
Ride-sharing on the daily home-work-home commute can help individuals save on gasoline and other car-related costs, while at the same time it can reduce traffic and pollution. This paper assesses the potential of ride-sharing for reducing traffic in a city, based on mobility data extracted from 3G Call Description Records (CDRs, for the cities of Barcelona and Madrid) and from Online Social Networks (Twitter, collected for the cities of New York and Los Angeles). We first analyze these data sets to understand mobility patterns, home and work locations, and social ties between users. We then develop an efficient algorithm for matching users with similar mobility patterns,  considering a range of constraints. The solution provides an upper bound to the potential reduction of cars in a city that can be achieved by ride-sharing. 

We use our framework to understand the effect of different constraints and city characteristics on this potential benefit. For example, our study shows that traffic in the city of Madrid can be reduced by 59\% if users are willing to share a ride with people who live and work within 1 km; if they can only accept a pick-up and drop-off delay up to 10 minutes,  this potential benefit drops to 24\%;  if drivers also pick up passengers along the way, this number increases to 53\%. If users are willing to ride only with people they know (``friends'' in the CDR and OSN data sets), the potential of ride-sharing becomes negligible; if they are willing to ride with friends of friends, the potential reduction is up to 31\%. 

\end{abstract}

\section{Introduction}
Ride-sharing is a promising approach for reducing the number of cars in a city, which is beneficial both for individuals \cite{owningCarCost} (reducing gasoline and other car costs) and for the city as a whole \cite{ameythesis} (reducing traffic and pollution). Carpooling lanes encourage commuters to share rides.  In recent years, a plethora of web and smartphone-based solutions have emerged for facilitating intelligent traffic management \cite{thiagarajan2009vtrack} and ride-sharing in particular. Early web-based systems, like {\tt carpooling.com}, and {\tt eRideShare.com}, provided matching of users  for long distance travel as well as for local and daily carpool and attracted a few million users across Europe and the US. More recently, companies like {\tt Avego.com, Lyft.com}, and {\tt Uber.com} provide smartphone apps that allow drivers and passengers to be matched; drivers make money, replacing traditional taxi services with a cheaper peer-to-peer solution.

Smartphone-based ride-sharing technology gains momentum but still needs to deal with several issues including safety  (traveling with strangers), liability (e.g., in case of accident), as well as the bootstrapping problem (the more users a particular ride-sharing service has, the more opportunities  to find users that can share a ride). 
However, even if/when the above technical problems were completely resolved (\eg assume that a particular ride-sharing service is adopted by everyone in a city), the success of ride-sharing would still depend on the underlying human mobility patterns and the layouts of a city, which ultimately determine the route overlap thus the opportunities for ride-sharing. 

In this paper, we seek to understand what is the potential decrease in the number of cars in a city  if people with similar mobility patterns were willing to use ride-sharing in their daily commute between home and work. This is clearly an upper bound to the actual benefit of any practical  system but it can be used to guide the deployment and policies regarding ride-sharing in a city. We assess this potential in four major cities in Europe and US (Madrid, Barcelona, New York, and Los Angeles) using mobile and social data sets. More specifically, two data sets are Call-Description Records (CDRs) provided by a major cell provider in Madrid and Barcelona, Spain. In addition, we collect two more data from Twitter (geo-tagged tweets) in New York and Los Angeles. A similar question has been asked before in \cite{tsao99}, where the authors, due to the lack of data, assumed a uniform distribution of home/work locations and concluded that ride-sharing has negligible potential. In contrast, we find  that ride-sharing can provide significant benefits, depending on the the spatial, temporal and social constraints for matching users, as well as the city and  data set used.

{\em Our contributions:}  We take the following steps.  

First, we infer {\em  home and work locations} for individual users from cell phone records and geo-tagged tweets, by adapting recent state-of-the-art techniques~\cite{Isaacman2011} to our setting. We also use the same (CDR and Twitter) data sets to obtain information about communication and explicitly stated ``friendship'' relations between users, which we later use to restrict rides between users that know each other or have common friends, in order to address concerns about riding with strangers.

Second, given a set of users with known home and work locations, we develop a {\em framework for matching} users that could share a ride so as to minimize the total  numbers of cars and provide rides to all users. We consider several constraints including: spatial (share rides only with people that are within a certain distance from their home and work location), temporal (share rides with people that want to depart/arrive within a time window from the desired departure/arrival time), and social (share rides only with people you know directly, or with whom you have common friends) constraints. We also consider two versions of the problem: \hwonly   -- ride-sharing between home and work locations, and \fullroute  -- allowing the possibility to pick up passengers along this route. Our framework is rooted at the Capacitated Facility Location Problem with Unsplittable Demand. Since this is an NP-hard problem~\cite{Korupolu1998},  and we have to match more than $272$K drivers and passengers, we develop efficient heuristic algorithms, namely  \hwmatching and \enroutematching to solve  the two aforementioned problems, respectively.

{\em Our results.}  We use our framework to assess the inherent potential of ride-sharing to exploit the overlap in people's commute in a city. We find that there is indeed significant potential for reducing traffic using-ride sharing, the exact magnitude of which depends on the constraints assumed for matching, as well as on the characteristics of the cities and the type of data set (CDR vs Twitter).  For example, our study shows that traffic in the city of Madrid can be reduced by 59\% if users are willing to share a ride with people who live and work within 1 km; if they can only accept a pick-up and drop-off delay up to 10 minutes,  this potential benefit drops to 24\%;  if drivers also pick up passengers along the way, this number increases to 53\%. If users are willing to ride only with people they know (``friends'' in the CDR and OSN data sets), the potential of ride-sharing becomes negligible; if they are willing to ride with friends of friends, the potential reduction is up to 31\%.  Albeit upper bounds to the actual benefit, these positive results (which  are in contrast with previous work, \eg \cite{tsao99}, that did not have all the information provided by our data sets) encourage the deployment and policies in favor of ride-sharing in urban settings.

The structure of the rest of the paper is as follows. Section 2 reviews related work. Section 3 presents our data sets, the methodology for inferring home and work location of individual users, and a characterization that provides insight into the next steps. Section 4 provides the formulation of the \hwonly problem, an efficient algorithm \hwmatching for solving it, and results from applying it on the data sets. Section 5 provides the formulation of the \fullroute problem, an efficient heuristic \enroutematching for solving it, and results from applying it on the data sets. In Section 6, we further restrict the matching and only allow users that know each other, or have common friends, to ride together. Section 7 provides a comparison across the four cities studied. Section 8 summarizes the results and concludes the paper.

\section{Related Work}

Traditionally, carpooling studies focused in characterizing the behavior of carpoolers, identifying the individuals who are most likely to carpool and explaining what are the main factors that affect their decision \cite{teal87}. The question we ask in this paper is city-wide: how much can ride-sharing reduce traffic? This has been studied before in \cite{tsao99}, but assuming a uniform distribution of home and work locations and concluding that ride-sharing has little potential for reducing congestion. In contrast, we infer home and work locations from CDR and Twitter data and we find that they are far from uniform.


Some carpooling systems have been designed based on GPS \cite{he2012mining, trasarti2011mining} data. He et al.  \cite{he2012mining} presents a frequency-based  route mining algorithm designed for GPS data and is used to extract frequent routes and recommend ride-sharing plans using the GPS traces of 178 individuals. Trasarti et al.\cite{trasarti2011mining} use GPS data to build the mobility profiles of 2107 individuals and match them based on the similarities of their mobility profiles; they also apply their algorithms to a GSM-like data set, which they synthesize by reducing the size of their GPS data set. To our knowledge, our work is the first attempt to study the potential of carpooling using CDR and OSN data. Although our data sets have less granular information in terms of user trajectories (essentially, we can only observe a user's location when she makes a call or posts a geo-tagged tweet), they have information about many more (orders of magnitude more) users than previous carpooling studies and thus are better positioned to answer the question about the city-wide benefit of ride-sharing.

Compared  to commercial ride-sharing systems, such as {\tt Avego, Lyft, Uber}: our work is partly based on publicly available (\eg geo-tagged tweets) as opposed to proprietary data;  it has a larger number of users for the cities studied; it takes into account social ties for matching drivers and passengers; and it assesses offline the city-wide benefit of ride sharing, as opposed to online matching of passengers with a small set of dedicated drivers.

Part of our methodology on inferring home and work locations for individuals uses and builds upon recent work by Isaacman et al. ~\cite{Isaacman2011, Isaacman2012} on identifying mobility patterns and important places from cell-phone data. The CDR call graph has been studied in~\cite{Cho2011}, \cite{eagle2009}. In this paper, we also use  CDR calls  or explicitly declared friendship in Twitter to infer whether two users know each other, and to restrict ride sharing accordingly.

There are related studies that focus on characterizing crowd mobility and urban environments using information from Twitter or Foursquare.  Wakamiya et al. \cite{wakamiya2011} and Fujisaka et al. \cite{fujisaka2010} have used geo-tagged Twitter datasets and its semantic content to study and characterize crowd mobility, and Frias-Martinez et al. \cite{noulas2012} for the characterization of land use. Foursquare has been used by Noulas et al. \cite{noulas2011exploiting}, \cite{noulas2013} for modeling crowd activity patterns. To the best of our knowledge, Twitter and Foursquare data have not been used for carpooling.

The most closely related work is the preliminary study by Cici et al.~\cite{coolestHotmobile2013}. Compared
to \cite{coolestHotmobile2013}, this UbiComp submission addresses the following additional issues: (i) using OSN data sets from Twitter (geo-tagged tweets for New York and Los Angeles) in addition to CDR data; (ii) including a CDR data set from Barcelona (in addition to Madrid); (iii) comparison between the four cities;  (iv) restricting ride sharing opportunities based on social ties; 
(v) estimating  the departure times from home/work from the actual data, as opposed to assuming a distribution.


\section{Extracting home and work from CDR and Twitter data}

The first step in assessing the benefits of ride-sharing is to identify where people live and work. In order to achieve this, we build on existing methodologies that have been proven to identify important locations in people's lives with adequate  accuracy \cite{Isaacman2011}. We apply their methodology with some modifications in order to make it applicable to our scenario.

\subsection{Data Sets}
We use four data sets in this study.  Two  call  description record (CDR) data sets were provided by a major mobile phone company in Europe for the cities of {\em Madrid} and {\em Barcelona}, Spain. Two data sets of geo-tagged tweets and Foursquare check-ins 
 were collected by us from the cities of {\em New York} and {\em Los Angeles}.

\subsubsection{Cell Phone Data}
Cell phone networks are built using a set of base  stations (BTS) that are in charge of communicating
cell phone devices with the network. Each BTS tower
has a geographical location typically expressed by its latitude and
longitude. The area covered by a BS tower is called a cell. The size of a cell varies from a few hundred square meters in an urban environment to up to 3 square kilometers in rural areas. 
At any given moment, one or more BTSs can give coverage
to a cell phone. Whenever an individual makes a phone call, the
call is routed through a BTS in the area of coverage. The BTS is
assigned depending on the network traffic and on the geographic
position of the individual.

Call Detail Records (CDRs) are  generated when a mobile phone makes or receives a phone call or uses a service (e.g., SMS, MMS, etc.).  Information regarding the time/date and the location of the BTS tower used for the communication are then recorded. More specifically, the main fields of each  CDR entry are the following: (1) the originating cellphone number (2) the destination cellphone number (3) the time and date that the call started (4) the duration of the call and (5) the BTS tower used by one, or both, cellphones. 
Note that no information about the exact position of a user within the area of coverage of a BTS tower is known. The CDR data set used for this study contains 820M calls from 4.7M mobile users in the metropolitan area of Madrid, and 465M calls from 2.98M users in the metropolitan area of Barcelona, during a 3-month period. For privacy reasons, no contract or demographic data was made available to us, and the originating and destination phone numbers were anonymized. 

\subsubsection{Twitter - Foursquare}
Many users access Twitter from mobile apps and some of them choose to include their current location (typically as GPS coordinates) in their tweets, thus making Twitter an important source of human mobility information. We used the Twitter's \textit{Streaming API} \cite{twitterStreamingApi} in order to obtain individual's mobility traces in large geographic areas. We collected geo-tagged  tweets from the metropolitan areas New York and Los Angeles for a period of four months -- from November 2012 until February 2013. This was possible thanks to Twitter's \textit{Public Stream Service} where you can specify the geographic area that you are interested in. As a result, we collected $3.23$M geo-tagged tweets from $155$K users in Los Angeles, and $5.70$M geo-tagged tweets from 225K users in New York. 

Twitter contains location information in geo-tagged tweets but lacks location semantics, which are crucial for identifying individual's home and work locations and commuting routes. We collected this information from Foursquare -- a large location-based social network with more than 30M users.
 Foursquare does not provide an API for data collection but Foursquare users can post their check-ins in Twitter and other OSNs. We obtained Foursquare check-ins from our Twitter data set: in  the geo-tagged tweets in Los Angeles we identified 134K  Foursquare check-ins from 13.6K users, and in New York we identified 362K Foursquare check-ins from 31.3K users. In addition,  we exploited another Foursquare data set with 1.47M check-ins from 40.1K users for training the algorithms to learns home and work locations.


\begin{figure}[t!]
\centering
\subfigure[Headquarters of Telefonica in Madrid]{\includegraphics[scale=0.80]{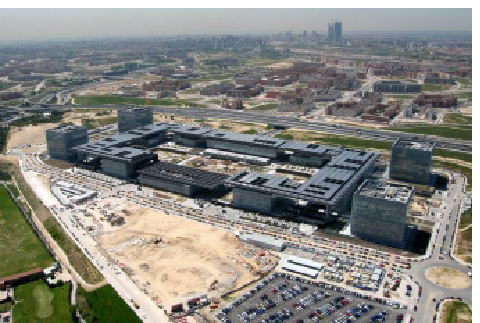}}
 \subfigure[Residential Area: Latitude:$40^o 30'13.45''$, Longitude:$3^o 38'07.69''$]{\includegraphics[scale=0.11]{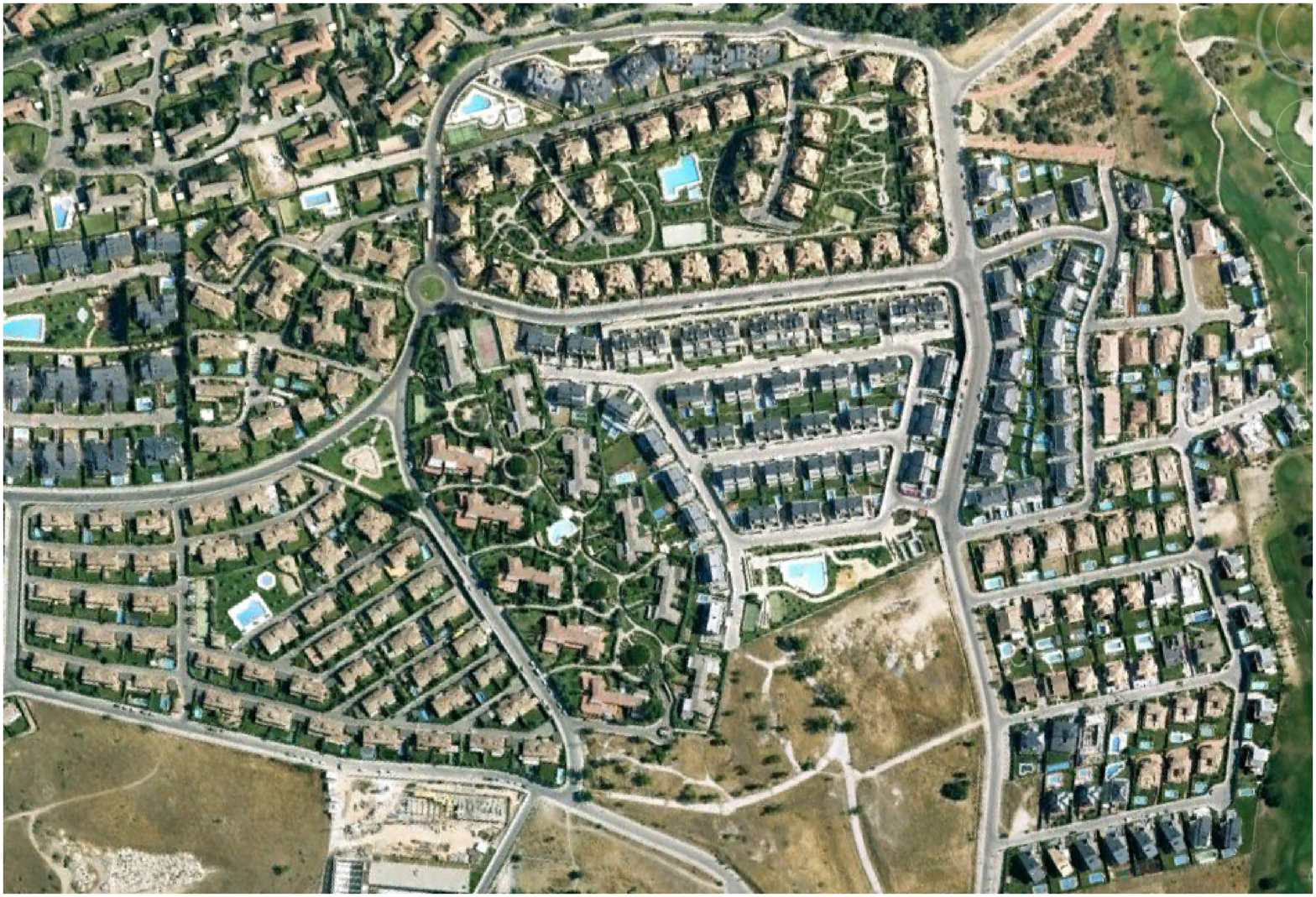}}
\caption{Example of strictly residential and strictly working areas}
\label{fig:hw-example}
\end{figure}

\begin{figure}[t!]
\centering
\subfigure[A ``ground truth'' user]{\includegraphics[scale=0.22]{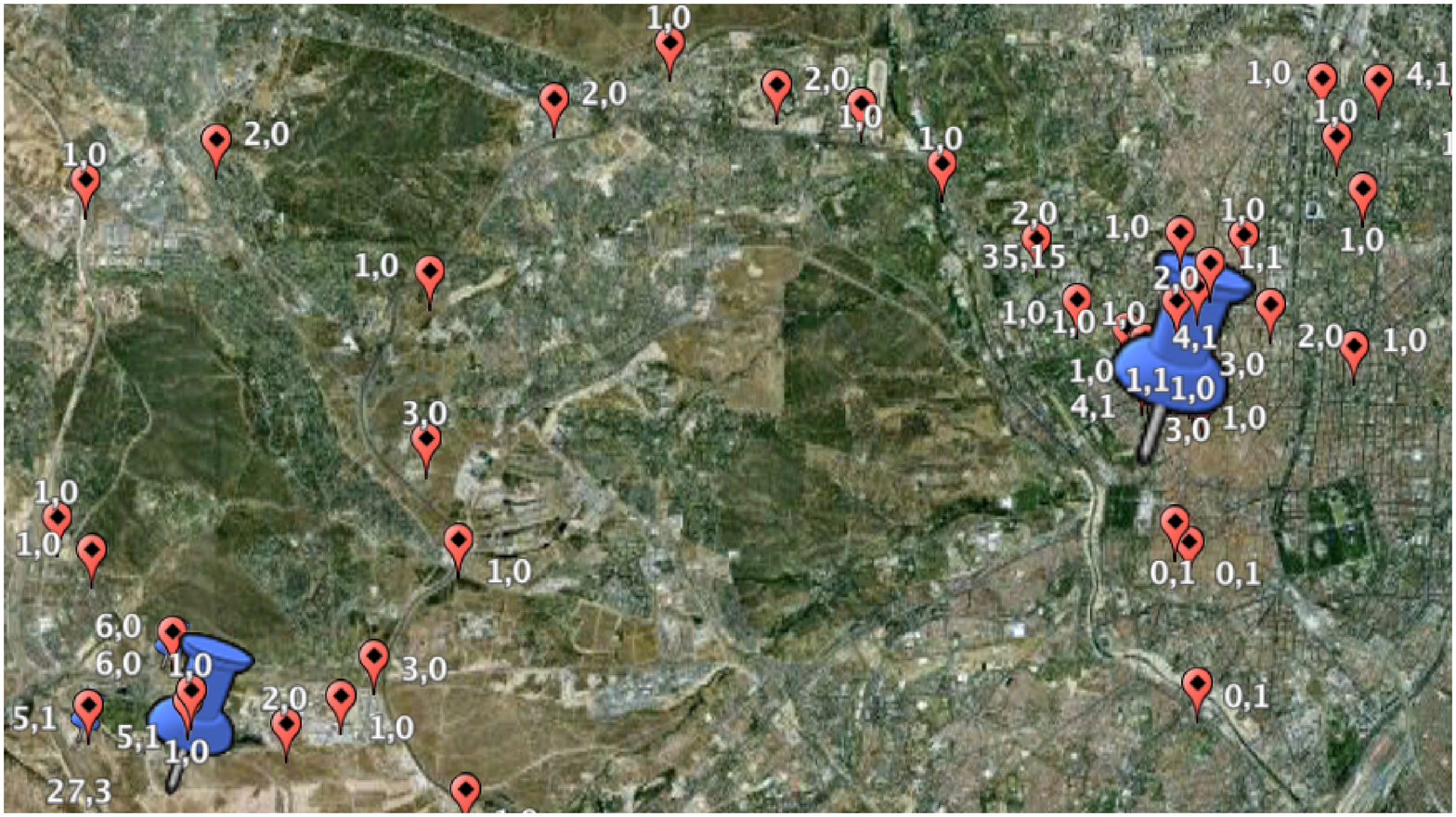}  \label{fig:gt-example-both}}
\subfigure[Zooming in at home]{\includegraphics[scale=0.22]{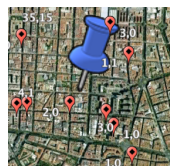} \label{fig:gt-example-home}}
\subfigure[Zooming in at work]{\includegraphics[scale=0.22]{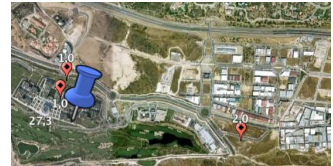} \label{fig:gt-example-work}}
\caption{The red paddles show the recorded cell towers for the user, while the blue pushpins the clusters. The white numbers next to each mark indicate the number of weekdays and the number of weekends the user appeared in that location. Also, the size of each mark is proportional to the number of days the user has appeared in that location.}
\label{fig:gt-example}
\end{figure}

\begin{table}[t!]
\caption {Users with identified Home/Work Locations}
\centering 
\small
\begin{tabular}{lccccc}
\hline 
City  & Source &  Number of users \\ \hline
Madrid  &  CDR &  272,479 \\
Barcelona  & CDR & 133,740 \\ \hline
New York  & Twitter & 71,977 \\ 
Los Angeles  & Twitter & 43,575 \\ \hline
\end{tabular}
\label{tab:identified}
\\Number of users with identified Home/Work locations.
\end{table}

\subsection{Identifying Home and Work}\label{subsec:homeident}

We apply the methodology of  Isaacman et al.~\cite{Isaacman2011} for identifying  important places for cell phone subscribers from (i) CDR data and (ii) ground truth for a subset of subscribers. First the recorded cell towers of a user are clustered to produce the list of places that the user visits.  
Then, regression analysis is applied on the ground truth users (their identified clusters and their true important locations) to determine the features that distinguish clusters that represent important places. Such features include: (1) the number of days that the user appeared on the cluster; (2) the duration of user appearances on the cluster; and (3) the rank of the cluster based on number of days appeared. Once important locations have been identified, the authors choose which of these locations is home and which is work. According to their results, the best features that characterize home and work are: (4) number of phone calls between  7PM - 7AM, i.e. \emph{Home Hour Events}, and (5) number of phone calls between 1PM - 5PM, referred to as \emph{Work Hour Events}.

For our CDR data set, first, we filter out users for whom we simply do not have enough data: \ie users with  less than 1 call per day on average, or less than 2 clusters with 3 days of appearance and 2 weeks of duration. Then, we tune the methodology of \cite{Isaacman2011} to our needs. More specifically, we build two classifiers, one for home and one for work, and we train them using the 5 features described above and the ground truth described in Sect.~\ref{sec:ground-truth}. Once the training is complete, we apply the classifiers to the rest of the users. Finally, after classification, we keep only the users who have only one location identified as home, and a different location identified as work, since we are interested only in commuters.  Applying the Home/Work identification methodology to our CDR data sets, we are able to infer the home and work locations of more than 270K individual users in the city of Madrid, and more than 133K users in the city of Barcelona (see Tab.~\ref{tab:identified}). 

For our Twitter-Foursquare data we apply the same methodology we applied in the CDR data set, with minor modifications. Since the Twitter-Foursquare data set is more accurate (fine-grained latitude-longitude coordinates of each check-in), we create much smaller clusters. Applying the methodology for identifying Home and Work  to our Twitter data sets, we are able to infer the home and work locations of more than 71K individual users in the city of New York, and more than 43K users in the city of Los Angeles Tab~ \ref{tab:identified}.

\subsection{Obtaining Ground Truth}\label{sec:ground-truth}
In~\cite{Isaacman2011}, a small set of 37 volunteers reported their most important locations, including home and work. This information was used to train the  classifiers that were applied to the remaining data set of around 170K mobile phone users. We did not have access to declared home and work locations from users, so we had to build our ground truth data sets.

\subsubsection{Ground Truth for CDR Data}

In the CDR case, we obtained our ground truth for a select subset of users based on our knowledge of the city of Madrid. In particular, due to its development pattern in the last 20 years, Madrid has many areas lying around its outer ring highways that are strictly residential and other ones that are strictly industrial. An example of the former are large residential development projects in previously isolated areas like the one depicted in Fig.~\ref{fig:hw-example}. Such areas offer a clear distinction between home and work and can be exploited to build our ``ground truth''. To this end, we selected $160$ users that appeared for many days in only one such residential area during 7PM - 7AM (assumed to be ``home'' hours), and only one such industrial area during 1PM - 5PM (assumed to be ``work'' hours).  Then, the location inside the residential area is pointed as the user's Home, while the location inside the industrial area is pointed as the user's work. 

For each one of the 160 users, we visually inspected their recorded locations through Google Earth. 
In Fig. ~\ref{fig:gt-example} we show one of these users: this individual lives in the location shown in Fig. ~\ref{fig:gt-example-home}, which is the top right blue marker of  Fig. ~\ref{fig:gt-example-both}, and work on the location shown in Fig. ~\ref{fig:gt-example-work}, which is the bottom left marker of Fig. ~\ref{fig:gt-example-both}. Note that home and work is defined as a cluster of cell towers, so home and work is an approximation of the real home/work locations.

\subsubsection{Ground Truth for Twitter data}
We used the Foursquare data to build the ground truth for the geo-tagged Twitter data sets by selecting users who appear more than seven days in a location tagged as \textit{Home(Private)}, and the same number of days in a location containing one of the tags: \textit{Professional, Office, or Work}. For each one of these users we define his home to be the location tagged as home with highest number of days of appearance, and as work the location tagged as work with most days of appearance. We also manually inspect their Twitter account and, when possible, their LinkedIn accounts as well. 

\begin{figure}[t!]
\centering
\includegraphics[scale=0.35]{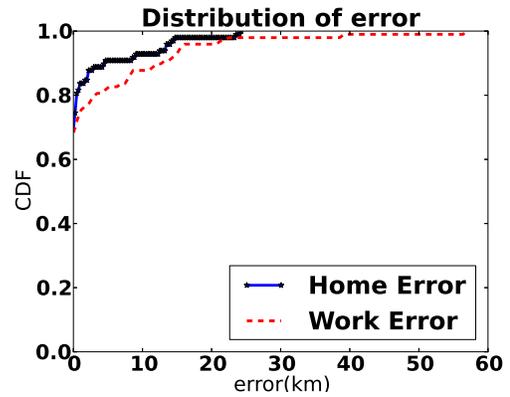}
\caption{Distribution of error for identification of h/w locations}
\label{fig:error_distr}
\end{figure}

\begin{table}[t!]
\caption {Comparison with related work}
\centering 
\small
\begin{tabular}{lccccc}
\hline 
Percentile & $25^{th}$ & $50^{th}$ & $75^{th}$ & $95^{th}$ \\ \hline
Our Home Error  & $0.0$&$0.01$&$0.49$&$13.62$ \\
Their Home Error  & $0.85$ & $1.45$ & $2.06$ & $6.21$ \\ \hline
Our Work Error  &$0.1$ &$0.03$ & $1.52$ & $16.09$ \\ 
Their Work Error  &$1.0$ &$1.34$ & $3.7$ & $34.17$ \\ \hline
\end{tabular}
\label{tab:compr}
\\Error in km. Comparing the home/work identification error to the related work.
\end{table}

\begin{figure}[t!]
\centering
   \includegraphics[scale=0.20]{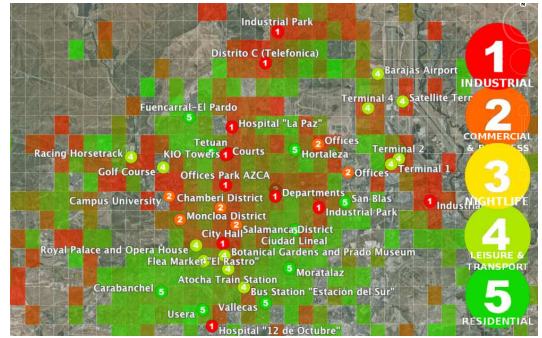}
   \caption{Characterizing Madrid based on our results}
   \label{fig:grid-madrid}
\end{figure}

\subsection{Validation}
Figure~\ref{fig:error_distr} shows the accuracy of the home/work identification methodology for our Twitter data set. In order to train the classifiers we used a previous Foursquare data set of $481$ users, and we did the tests in a group of $98$ ground truth users for whom we have both geo-tagged tweets and Foursquare check-ins. For these users, we found their home and work addresses using their geo-tagged tweets and then we compared the identified home and work locations with the real ones. In Tab. ~\ref{tab:compr} we compare the accuracy of the home/work identification methodology with the reported accuracy in \cite{Isaacman2011}. We see that in the case of the $75^{th}$ percentile the home error has decreased by 76\% , and the work error has decreased by 59\%. For a small number of cases, our error is higher. We attribute our overall higher accuracy to the more precise location information in the Twitter-Foursquare data sets. 

In Fig.~\ref{fig:grid-madrid} we break the city of Madrid into a grid and color each square of the grid with a combination of green and red. If the number of inferred home locations is higher than the number of work locations, then the color of the square is closer to green, otherwise it is closer to red.  We use an existing study~\cite{Soto2011} to obtain a characterization of locations in Madrid (industrial, commercial, nightlife, leisure and transport, residential).  We annotate such areas in Fig.\ref{fig:grid-madrid} using numbered circles, \eg, the headquarters of Telefonica in Madrid is one of the two red circles on the top of the figure. We observe that the squares that we colored red contain indeed more circles indicating industrial and commercial zones, than residential zones. Also, squares colored green contain more residential zones than industrial zones.

\begin{figure}[t!]
\centering
\subfigure[Measured]{\includegraphics[scale=0.153]{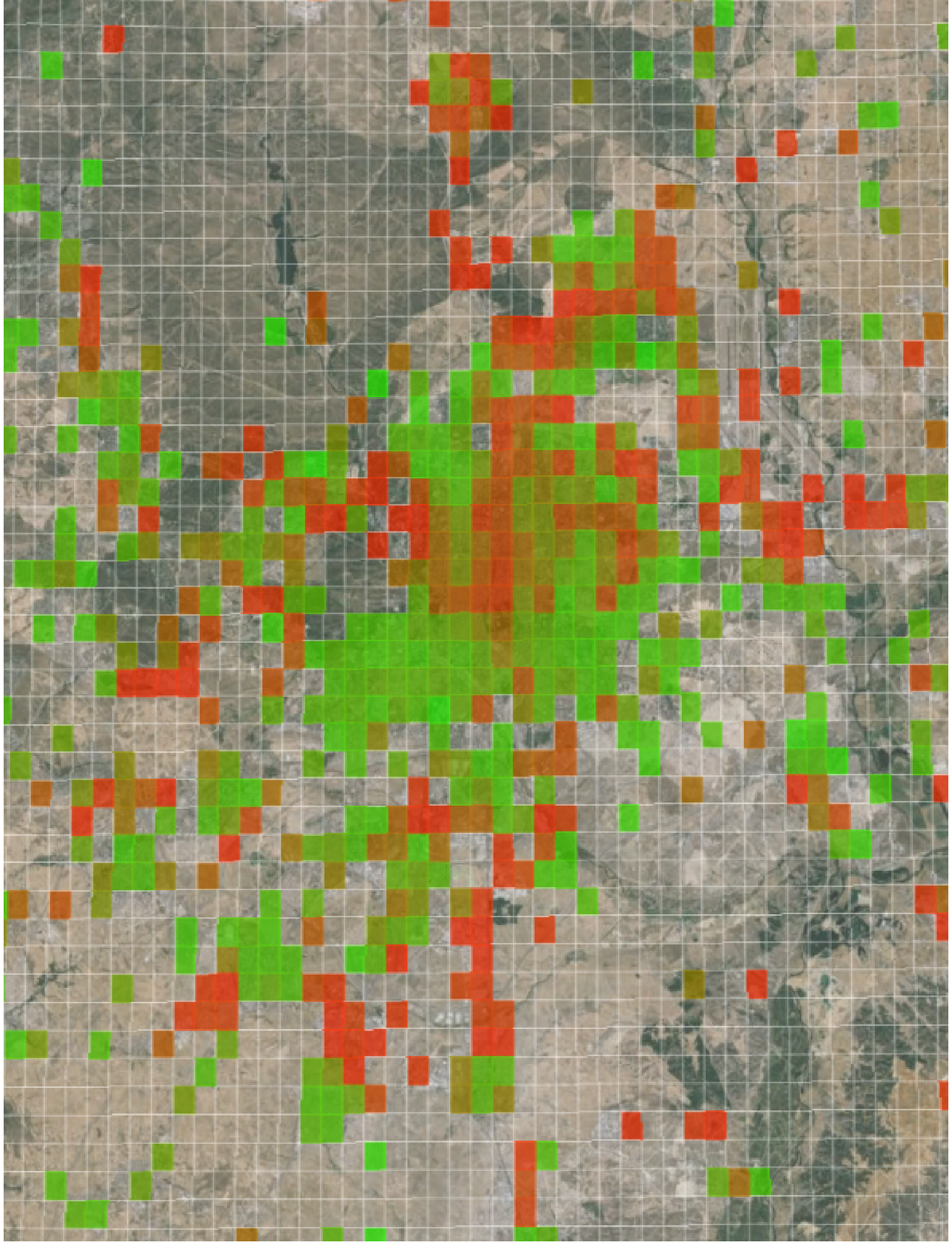}\label{fig:hw_estimated}}
 \subfigure[Uniform]{\includegraphics[scale=0.160]{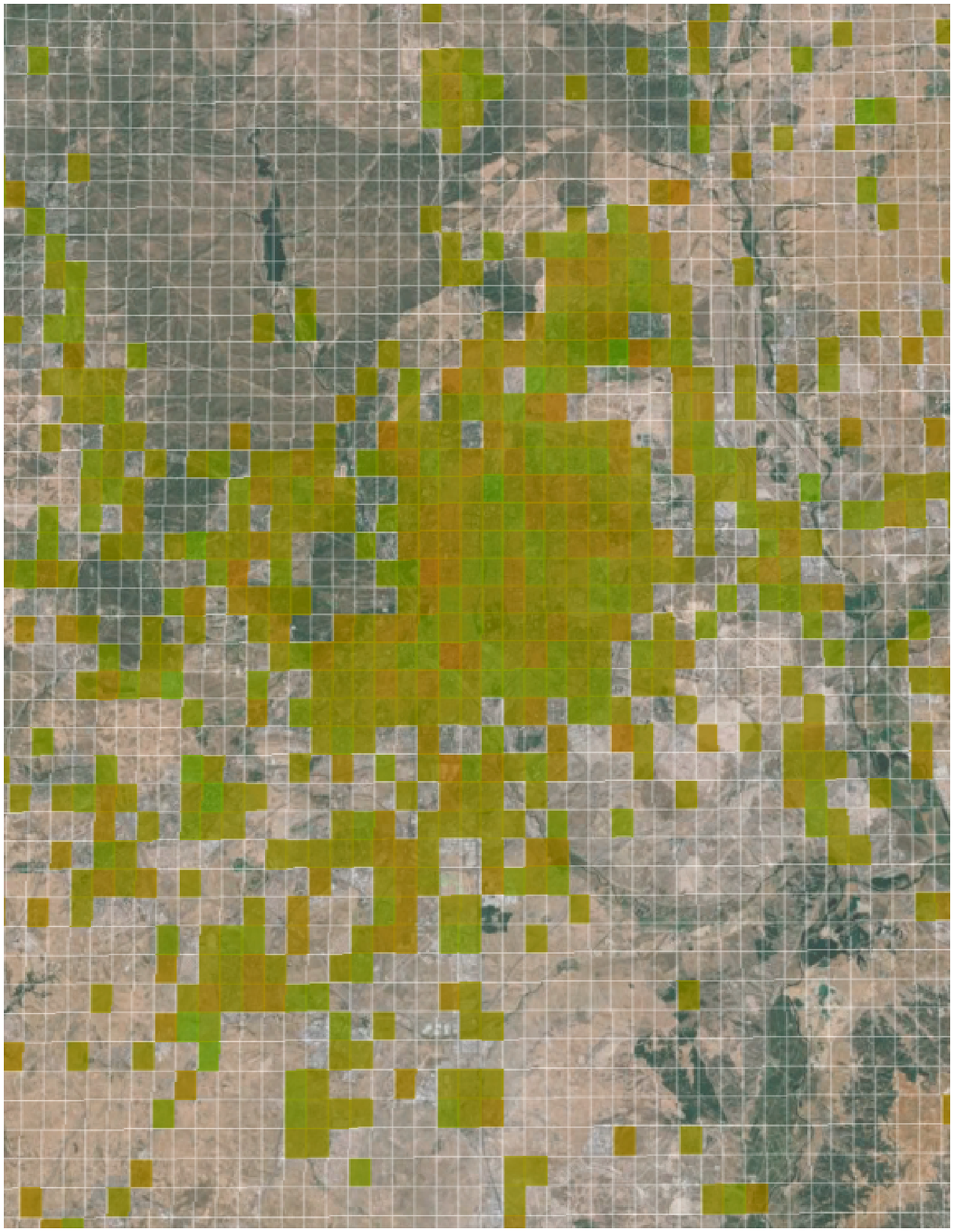}\label{fig:hw_uniform}}
\caption{Home/Work distribution comparison between Measured (from the data) and Uniform. Red is for works , and green for homes.}
\label{fig:estimated_vs_uniform}
\end{figure}

\begin{figure}[t!]
\centering
\includegraphics[scale=0.30]{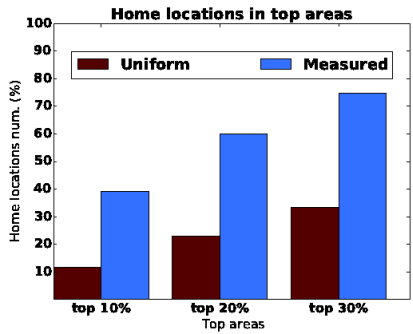}
\caption{\# locations (home) for top 10\%, 20\%, and 30\%  of the areas. }
\label{fig:estimated_vs_uniform2}
\end{figure}

\begin{figure}[t!]
\centering
\subfigure[Top home location]{\includegraphics[scale=0.2]{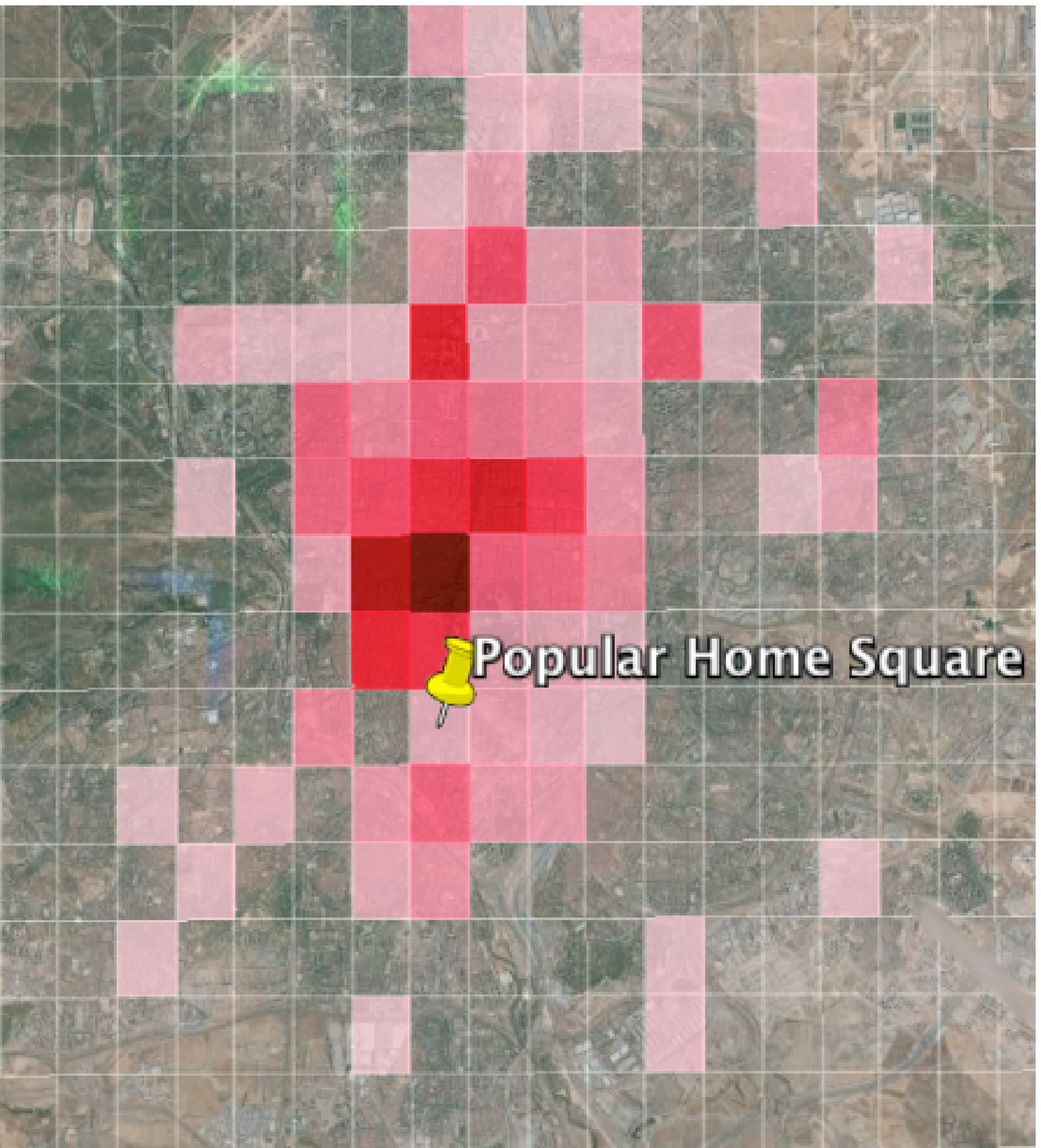}\label{fig:top_home}}
 \subfigure[CDF of home-work distance]{\includegraphics[scale=0.18]{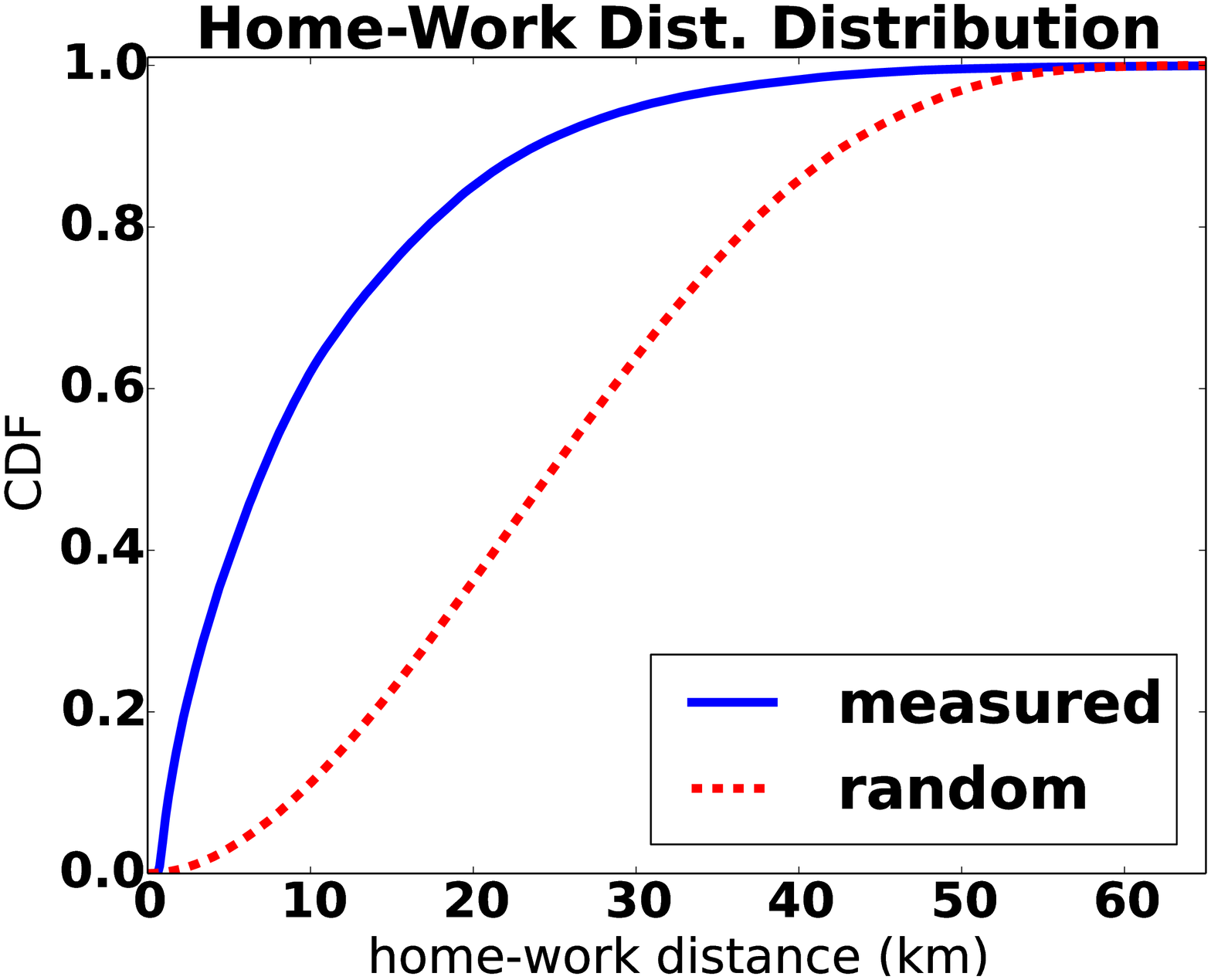}\label{fig:home_work_dist_cdf}}
\caption{Distance between home and work locations. \ref{fig:top_home} shows the square grid with most homes (yellow paddle), and where are the corresponding 
work locations; stronger the colors indicate higher concentration of work locations. }
\label{fig:home_work_distance}
\end{figure}

\begin{figure}[t!]
\centering
   \includegraphics[scale=0.23]{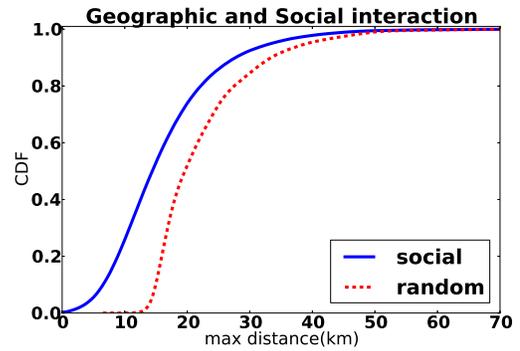}
   \caption{Geographic distances between users who have social ties -- inferred from the calls -- and random strangers. The distance between two users $u$ and $v$ is the maximum of their home and work distance.}
   \label{fig:social_and_geo}
\end{figure}


\subsection{Differences from the Uniform Distribution}

We find that the distributions of the home and work distribution are far from uniform, which was assumed in prior work \cite{tsao99}:

{\em Segregation of residential and working areas:} According to Fig.~\ref{fig:estimated_vs_uniform},  Madrid contains residential and working (office or industrial) areas. In industrial areas (shown in red) there is relatively large number of working places, while in residential areas there is a relatively large number of homes, Fig.~\ref{fig:hw_estimated}. To illustrate the difference, we show how the city would look if the home/work distribution were uniform, Fig.~\ref{fig:hw_uniform}

{\em Different density:} The density of home and work locations in various areas is quite different from  uniform. Fig.~\ref{fig:estimated_vs_uniform2}. 30\% of top home areas -- areas with most home locations -- contain 75\% of the homes; if home/work distribution was uniform then the top 30\% of home areas would contain only 30\% of the homes.

{\em Relatively short home-work distances:} As seen in Fig.~\ref{fig:home_work_distance}, users tend to work close to where they live. For the grid square with the highest number of users who have their home there, Fig.~\ref{fig:top_home}, the corresponding work locations tend to be close by. Also, according to Fig. 7(b), the home-work distances are shorter compared to what they would be if  home and work were randomly distributed.  

{\em Geographic distances and social ties:} In Section ~\ref{sec:social}, we will consider social ties between between users (as indicated by  call in the CDR data or by an explicitly declared "follower" and "followee" in Twitter). in  Fig.~\ref{fig:social_and_geo}, we compare the average geographic distance from each user $u$  to her friends, versus her geographic distance to randomly selected strangers (i.e., users who are not neighbors of $u$ in the CDR or Twitter social graph).  According to Fig.~\ref{fig:social_and_geo} the  geographic distance between users who have social ties are shorter, on average,  than the geographic distances between strangers.

\begin{figure}[t!]
\centering
\includegraphics[scale=0.23]{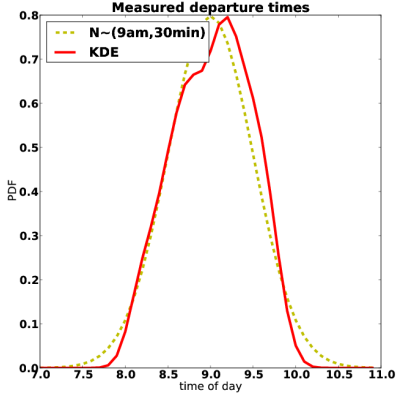}
\caption{Distribution of Departure Times from Home. The Dotted line is a Gaussian distribution with mean at 9 am, and standard deviation 30 minutes. The red continuous line is what 
we get via Kernel Density Estimation from our data.}
\label{fig:departure_times}
\end{figure}


\subsection{Departure Times}\label{sec:depart_times}


We estimate departure times from consecutive home/work calls. More specifically, we use pairs of calls where  one of them is a home call, the other a work call, and the time difference between the calls is less than $2*$trip{\_}time, where trip{\_}time is the time distance between home and work, as obtained from a popular Online Map service.

For each user, we find her departure time from home by taking the median of the calls, that: 1) were made between $8$ am and $10$ am from home, and 2)  were followed by a work call no more than $2*$trip{\_}time later. Similarly, we find her work departure time, by taking the median of the calls, that : (i) were made from work between 4pm and 6pm, and (ii) were followed by a home call no more than $2*$trip{\_}time later.

The distribution of home departure times for all individuals who had such calls -- each individual is required to have at least three such calls is shown in Fig.~\ref{fig:departure_times}; there were $484$ such users in our data set. The departure time from work follows a similar distribution, which is omitted due to lack of space.


\section{End-points Ride-sharing}

Here we formulate the problem of  \hwonly\ \ie, ride-sharing among people that live in a common area and work in another common area.  We develop a practical algorithm, apply it to the CDR and Twitter data sets used to estimate home/work locations, and compute the number of cars that can be reduced under different scenarios.

\subsection{Formulation}

Let $V$ denote a set of potential drivers and $c(v)$ the capacity, in terms of available seats, of the car of driver $v\in V$ and $p(v)$ a penalty paid if driver $v$ is selected for driving his car and picking up passengers. Let $h(v,u)$ denote the geographic distance between the home locations of drivers $v$ and $u$ and $w(v,u)$ the corresponding distance between their work locations. Let $\delta$ denote the maximum distance between a driver's home/work and the home/work of passengers that he can pick up in his car, \ie, $v$ can have $u$ as passenger only if: $\quad max(h(u,v), w(u,v)) \leq \delta$

Let $d(v,u)$ denote a virtual distance between $v$ and $u$ defined as follows:
\[d(v,u) = \left \{ \begin{array}{l}
h(v,u)+w(v,u), \\
\textrm{if}\quad \max(h(v,u),w(v,u))\leq \delta\\[1em]
\infty, \quad \textrm{otherwise} 
\end{array} \right. \]

Our objective is to select a subset of drivers $S\subseteq V$, and find an assignment $a : V \rightarrow S $, that minimizes $P(S)+D(S)$, the sum of penalty and distance costs, while satisfying the capacity constraints of cars. The two costs are defined as follows:
\[ P(S)=\sum_{v\in S} p(v) \quad \mbox{and} \quad D(S)=\sum_{v\in V} d(a(v),v) \]
where $a(v)\in S$ is the driver in $S$ that is assigned to pick up passenger $v$ (can be himself if $v$ is selected as a driver). By setting $p(v)>2 \delta \cdot c(v)$ we can guarantee that an optimal solution will never increase the number of cars used in order to decrease the (pickup) distance cost between a driver and its passengers.  The above problem is an NP-hard \emph{Capacitated Facility Location Problem with Unsplittable Demand} in metric distance: the set of potential drivers corresponds to the set of locations; the set of chosen drivers corresponds to opened facilities; car capacity corresponds to facility capacity; distance $d(v,u) $ corresponds to the cost of assigning a location $v$ to the facility $u$. Efficient approximation algorithms are known for this type of facility location problem~\cite{Korupolu1998}.

\begin{figure}[t!]
\centering
\subfigure[ When $\sigma = 10$ ]{\includegraphics[scale=0.35]{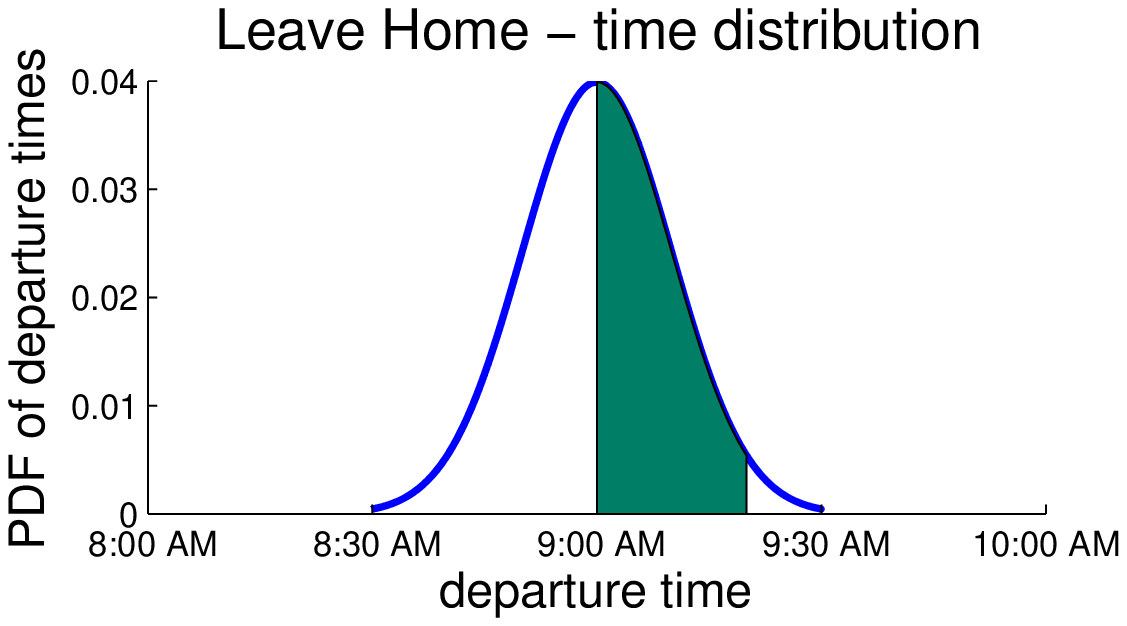}}
 \subfigure[ When $\sigma = 20$]{\includegraphics[scale=0.35]{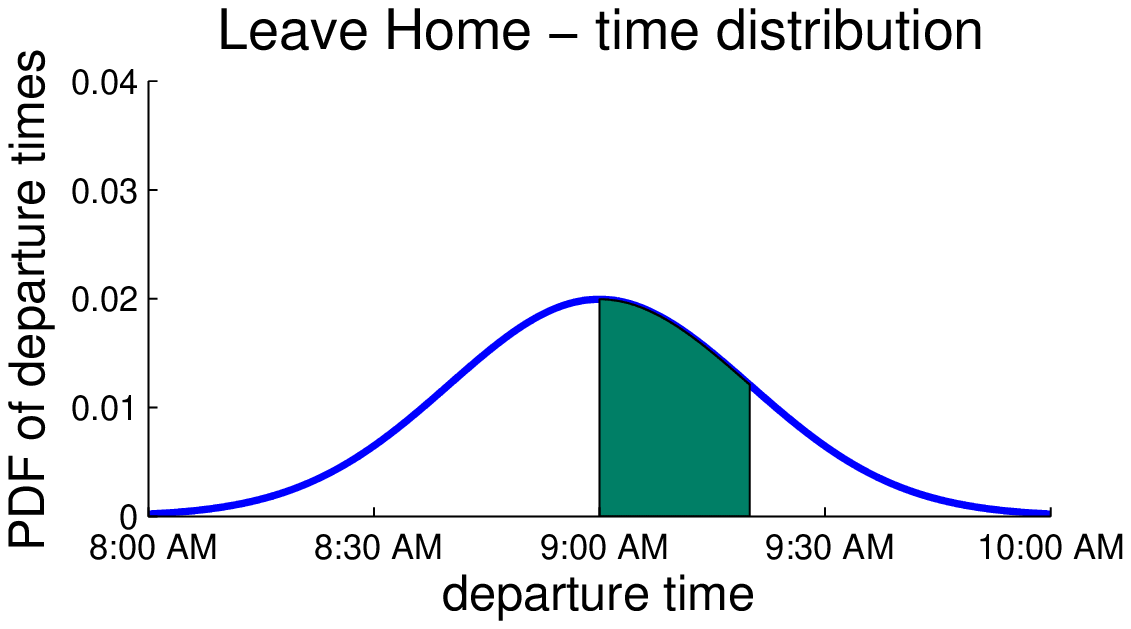}}
\caption{Assuming users $u$ leaves home at 9:10, the users departing with 10 min difference are in the green area under the curve.}
\label{fig:departure-example}
\end{figure}

The above formulation finds the minimum number of cars needed when there are no timing constraints around departure and return times from home and work. Next we refine the formulation to include time. We assume that departures from home and work follow Gaussian distributions, centered at 9 am and 5 pm respectively, with standard deviation $\sigma$ (see Sec.~\ref{sec:depart_times}). Also, we introduce the wait tolerance $\tau$ that captures the maximum amount of time that an individual can deviate from his normal schedule in order to share a ride, Fig.~\ref{fig:departure-example}. More specifically, if $LH(u)$ expresses the time a person $u$ leaves home to go to work, and $LW(u)$ expresses the time she leaves work in order to return to home. Then, two people $u$ and $v$, can share a ride only if: 
 \[ max(| LH(u) - LH(v)|, |LW(u) - LW(v)|)  \leq \tau \]
The introduction of the temporal constrains will only change the virtual distance between $v$ and $u$ :
\[d(v,u) = \left \{ \begin{array}{l}
h(v,u)+w(v,u), \\
\textrm{if}\quad \max(h(v,u),w(v,u))\leq \delta \\
\textrm{AND}\quad |LH(u) - LH(v)| \leq \tau \\
\textrm{AND}\quad |LW(u) - LW(v)| \leq \tau \\[1em]
\infty, \quad \textrm{otherwise} 
\end{array} \right. \]

\subsection{A Practical Algorithm}\label{sec:practical}

In this section we show how to modify the existing approximation algorithm~\cite{Korupolu1998} for the facility location problem described above and obtain a faster heuristic that can cope with the size of our data set.  

The existing algorithm starts with an initial random solution and improves it iteratively via local search.
At each iteration there are $O(n^2)$ candidate solutions, where $n$ corresponds to the number of potential drivers, and for each one of them we find the assignment (passengers to drivers) that will minimize the cost; this can be done in polynomial time by solving an appropriately defined instance of the \emph{transportation problem}.  The algorithm terminates when local search cannot find a better solution. 

We modify the algorithm in three ways. First, since the quality of the solution depends mostly on the number of drivers, we try to keep that number as low as possible. Therefore, we use the b-matching \cite{Cechlarova2005} algorithm to generate the initial solution, instead of generating it randomly. The input to the b-matching algorithm consists of the set of potential drivers $V$, a function $p(v)$ that defines the
set options for a potential driver $v$ {\it i.e. } $p(v) = \{u | d(u,v) < \inf \}$, and a global ordering of the potential drivers, $O$. The global ordering will be based on the number of options; the fewer the options, the higher the position in $O$. By using b-matching with a global order we are guaranteed to find a solution in $O(n)$ time \cite{Cechlarova2005}. For each match generated by b-matching, we assign the potential driver with the most occupied seats to drive; we make sure that every user in $V$ appears in only one car. This solution proves much better than the random one by paying $O(nlog(n))$ for sorting the users to generate the global preference list and $O(n)$ for the matching.

Second, solving a transportation problem with 270K users is hard. Therefore, we need to modify the local search steps of the approximate algorithm. Given an initial solution we leave the users commuting in cars of four as they are and search for better assignments only for the rest. This way the size of the transportation problem will be reduced and that would speed up the process of generating the assignment.

Third, reducing the size of the transportation problem is not enough; we also need to reduce the neighborhood of candidate solutions. Given an initial set of drivers, $S$, we create a fixed size neighborhood, where each solution $S'$ is created by doing random changes in S. The reason why we do that is because considering all potential solutions that differ from S only by one, means that we have to examine $O(n^2)$ candidate solutions; that makes each iteration very expensive. Therefore, the fixed size solution helps us speed up the time we spend in each improvement step. 

Without the above modifications it would be impossible to solve the problem in real time. Solving an instance of the transportation problem for 270K users required a couple of hours for $\delta = 0.6$ km, and even more when $\delta = 0.8$ or $\delta=1.0$ km. Therefore, solving $O(n^2)$ such problems for a single iteration becomes too time consuming. Moreover, most of the time the solution of the b-matching algorithm was so good that the gain from the improvement steps would be insignificant.

\begin{figure}[t!]
\centering
\includegraphics[scale=0.32]{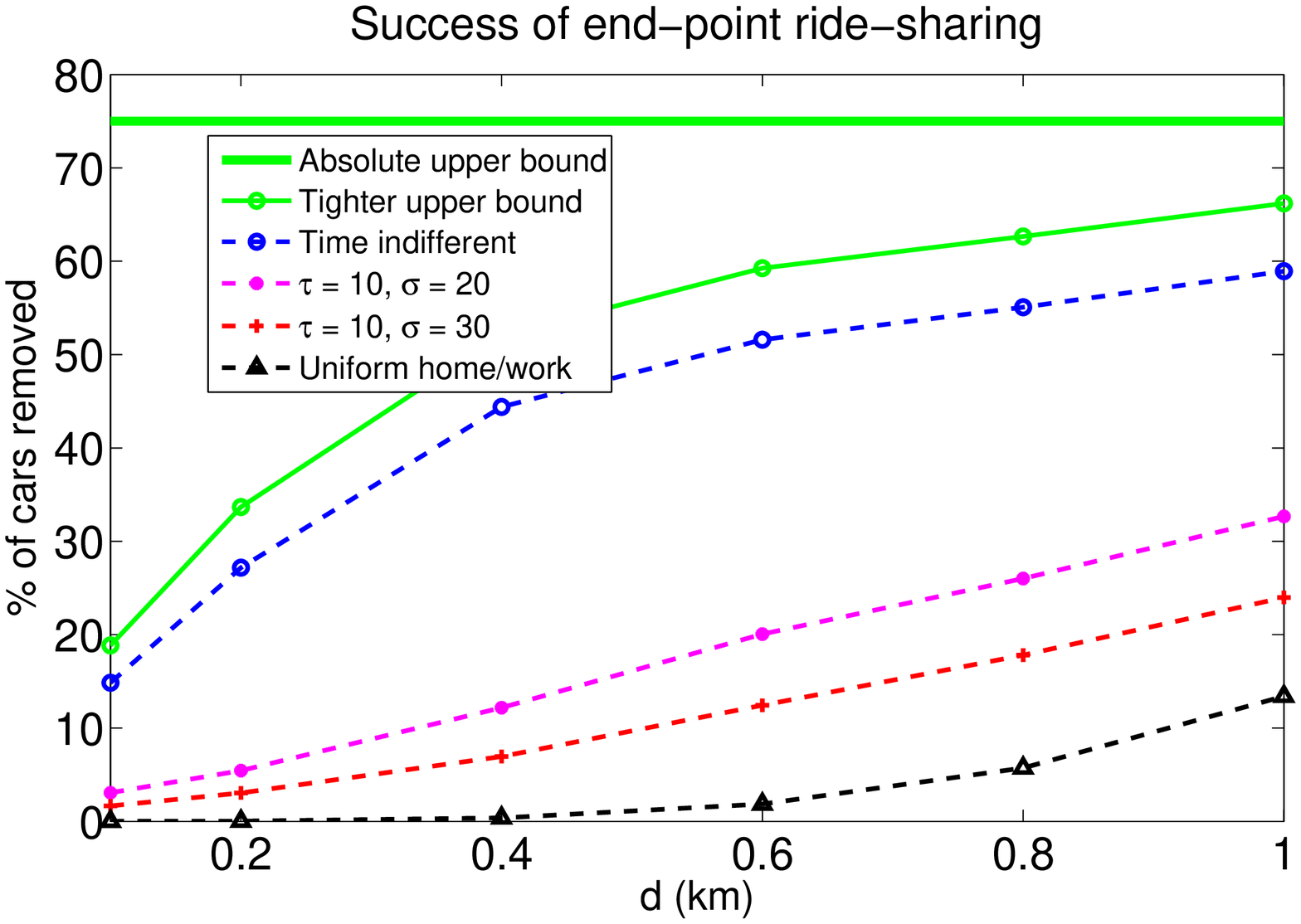}
\caption{Benefits from \hwonly.}
\label{fig:results}
\end{figure}

\begin{figure}[t!]
\centering
   \includegraphics[scale=0.20]{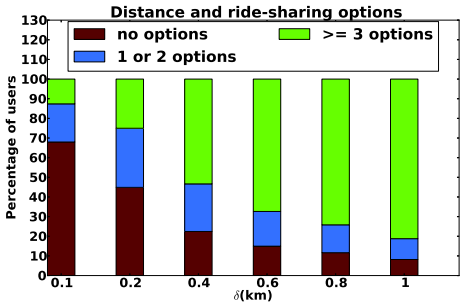}
   \caption{How $\delta$ affects the ride-sharing options}
   \label{fig:explanation}
\end{figure}

\subsection{Results}

A this point we are ready to calculate the effectiveness of \hwonly~ based on our data sets. For ease of exposition we will focus on the Madrid metropolitan area (we cover the other three cities in Sect.~\ref{sec:allcities}). We reduce the size of our data set by randomly selecting only 60\% of the users.
We do that to capture the fact that only 60\% of the population has a car in the area of Madrid \cite{carOwnershipMadrid}. We also show results for the case that half of the car owners use their car at their daily commute (the results are quantitatively close). For the remaining of the section, we will refer to users who can share rides with a specific user $v$, as \emph{options of $v$}. Subsequently we compute the percentage-wise reduction of cars 
\[  \textrm{success} = \frac{\textrm{ \#(init. cars)} -  \textrm{\#(ride-sharing cars)} }{\textrm{\#(init. cars)}} \cdot 100  \]
using the following algorithms:

{\em Absolute upper bound:} Given our definition of success, we cannot do better than 75\%. This is the case when all cars carry 4 people.

{\em Tighter upper bound:} All users with at least one ride-sharing option, are assumed to commute in cars of 4.  

{\em Time-indifferent matching ($\tau=\infty$):} This is the practical algorithm described in Sect.~\ref{sec:practical}

{\em Time-aware matching:} This is the version of the algorithm that considers timing constraints under the assumption of normally distributed departure times.

{\em Uniform home/work:} The potential of ride-sharing would be if home/work locations were distributed uniformly. 

Fig.~\ref{fig:results} presents what happens when the drivers are willing to tolerate a detour of $\delta$ and deviate $\tau$ minutes from their departure times, in order to share the same car with another individual. The results show that with even modest delay tolerance of 10 minutes and detour distance of 1.0 km (a couple of city blocks) more than 20\% of the cars can be saved. This is more than half of the absolutely optimal performance. Increasing either of the two parameters improves the success ratio.  The diminishing improvement with increasing $\delta$ can be explained by the number of  options users have given the distance $\delta$. In Fig.~\ref{fig:explanation} the red color represents the users with no options, the blue color the users with 1 or 2 options, and the green color the users with 3 or more options. We can see that the success of ride-sharing is proportional to the number of users who have 3 or more options.

Also, in Fig.~\ref{fig:results} we can see that the potential of \hwonly is quite small in the case of uniformly distributed home/work locations; note that no time constraints were applied in this case. If we apply time constrains too, then the success of \hwonly is even smaller, e.g. for $\delta = 1$ km, $\tau=10$ min, and $\sigma=30$ min, its potential becomes 0.2\%

\section{En-route ride-sharing}

The effectiveness of ride-sharing can be greatly enhanced by picking up additional passengers en-route. For example a driver that lives in a sparsely populated area might not have any neighbors to fill his seats but once he enters the city he might be able to pick several passengers that have routes ``covered'' by his own. Focusing in our Madrid example, and in order to quantify the benefits of en-route ride-sharing we obtain routes from Google Maps for the 270K users of the CDR data set and extend the algorithm of Section~\ref{sec:practical}.


\subsection{En-Route Algorithm}

We use an iterative algorithm with the following steps in each iteration.
\begin{enumerate}
\item Run the basic \hwonly~ algorithm.
\item Exclude from the solution cars that get fully packed (a car of 4). Then order cars in decreasing order of passengers and start ``routing'' them across the urban environment (Madrid in our running example) using data from Google maps. 
\item When the currently routed car $v$ meets a yet un-routed car $v'$, then $v$ is allowed to steal passengers from $v'$ as long as it has more passengers than $v'$ (a rich-get-richer strategy). Whenever a routed car gets fully packed it is removed from further consideration. Whenever a car with a single passenger is encountered the number of cars is reduced by one.
\end{enumerate}
These steps are repeated until there is no possible improvement.  It can be shown (omitted for lack of space) that the rich-get-richer rule leads to convergence.

\begin{figure}[t!]
\centering
\includegraphics[scale=0.32]{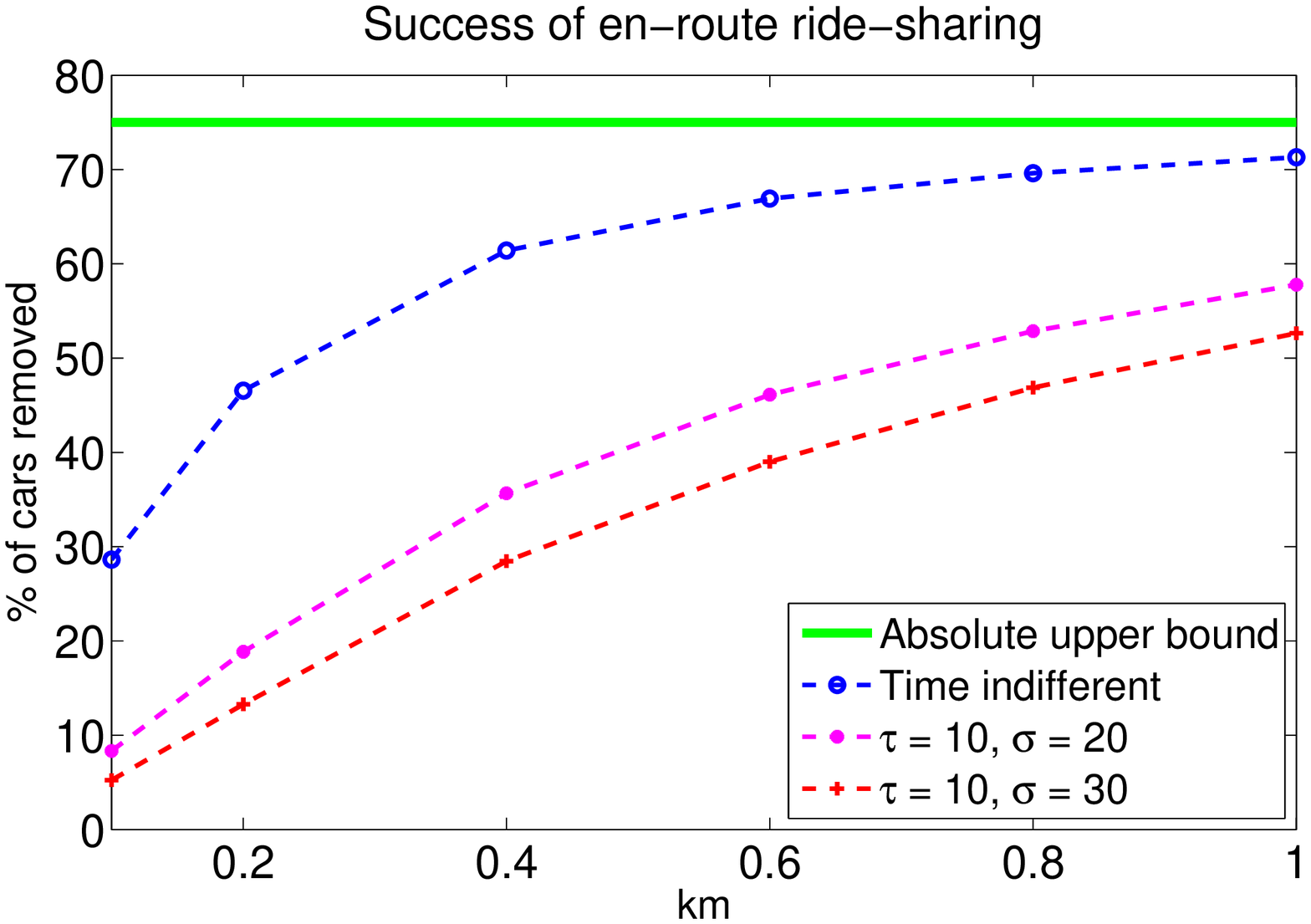}
\caption{Benefits from \fullroute.}
\label{fig:results-pick-ups}
\end{figure}

\begin{figure}[t!]
\centering
\includegraphics[scale=0.35]{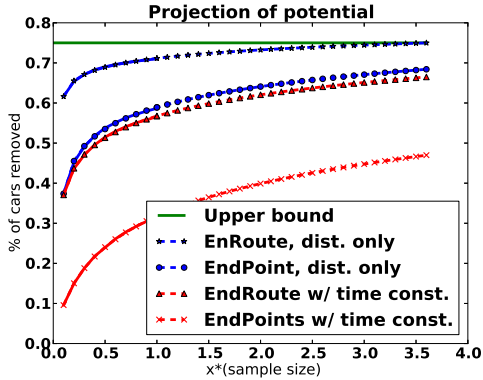}
\caption{Extrapolation to commuters' size. ``Sample" refers to the entire 270K location data set of Madrid. The solid lines correspond to values generated from our data set, while the dashed lines correspond to values generated through extrapolation.}
\label{fig:projection}
\end{figure}

\begin{table}[t!]
\caption {Summary of results for Madrid, $\delta = 1.0$ km}
\small
\begin{tabular}{lccccc}
\hline 
Sample &$\delta$& $\tau$&$\sigma$&\hwonly&\fullroute \\ 
(\%) & (km) &  (min)  & (min) & (\%) & (\%) \\ \hline
30 & 1.0  &  --  & --  &  54 & 65  \\
30 & 1.0  &  10  & 30 & 17 & 47 \\ \hline

60 & 1.0  &  --  & --  & 59 & 70 \\
60 & 1.0  &  10  & 30 & 24 & 53 \\ \hline

100 & 1.0 &  --  & --  & 62 & 71 \\
100 & 1.0 &  10  & 30 & 30 & 56  \\ \hline

360 & 1.0 &  --  & --  & 70 & 75 \\
360 & 1.0 &  10  &  30 & 44 & 65 \\ \hline

\end{tabular}
\label{tab:summary}

This table shows how the population size affects the performance of \hwonly and
\fullroute. ``Sample" refers to the entire 270K location data set of Madrid. 100\% means using all of it. 30\% and 60\% means using  subsets of it. 360\% means projecting the potential to the entire commute population of Madrid ($\times$3.6) as explained in Sect.~\ref{subsec:enrouteresults}.
\end{table}

\subsection{Results}\label{subsec:enrouteresults}

Fig.~\ref{fig:results-pick-ups} shows the performance of \fullroute. To make the comparison with \hwonly~ easier we summarize results from both approaches in Table~\ref{tab:summary}. By comparison, it is possible to verify the significant improvement obtained through \fullroute, which in several cases comes within 10\% of the optimal performance.

\vspace{2pt}

\noindent {\bf Projection to the entire commute population:} All previous results have been produced based on the 270K subscribers in Madrid for which we were able to obtain a credible estimation of Home and Work following the methodology of Sect.~\ref{subsec:homeident}. This, however, represents only roughly 8\% of the total population of the city. To get a feeling of the ride-sharing potential based on the entire population, for which we do not have location information, we employ a simple projection method. We repeat the calculation of ride-sharing with different subsets of our total 270K users and plot the empirical function connecting the sample size with the resulting ride-sharing potential. Then we fit numerically these data points to the best analytic function that describes them and evaluate this function at the desired population point that correspond to the entire city. The results are depicted in Fig.~\ref{fig:projection} and summarized in Tab.~\ref{tab:summary}. We can see from this plot that the population size has a progressively diminishing results on the ride-sharing potential. In the remainder of the article we will report results for both our 8\% sample and projected results to the entire commuting population.

\section{Social Filtering - Riding with Friends of Friends}\label{sec:social}

\begin{table}[t!]
\caption {Graph sizes}
\begin{tabular}{lcccc}
\hline 
\centering
Graph & Nodes & Edges  & $\textrm{Mean} \atop \textrm{degree}$   & $\textrm{Median}  \atop \textrm{degree}$\\
       & \#  &  \#  &   &  \\ \hline
call graph Madrid & 4M &  21M  & 6.0 & 1 \\ 
twitter graph NY &  132K & 725K & 10.95  & 5   \\
\end{tabular}
\label{tab:our-graphs}\\
\end{table}

\begin{table}[t!]
\caption {Social Filtering}
\small
\begin{tabular}{lcccc}
\hline 
city & filter &\hwonly&\fullroute & $\textrm{\fullroute} \atop \textrm{extrapolation}$ \\ 
 & & (\%) & (\%) & (\%) \\ \hline

Madrid & no filter & 30 & 56 & 65  \\
Madrid & 1-hop & 0.26 &  1.1 &   --   \\
Madrid & 2-hop & 3.7 & 19 & 31  \\ \hline
NY & no filter & 20 & 44 & 68 \\
NY & 1-hop  & 0.18 & 1.2 & -- \\
NY & 2-hop & 2.1 & 8.2 & 26  \\ \hline
\end{tabular}
\label{tab:soc-constr-data}
The potential or \hwonly and \fullroute for $\delta = 1.0$ km (distance constr.), $\tau$=10, $\sigma$ = 30 (time constr.). The third and the forth column show the potential for sample size, while the last column shows the potential of ride-sharing extrapolated to the commuters' population.
\end{table}

\begin{figure}[t!]
\centering
\includegraphics[scale=0.35]{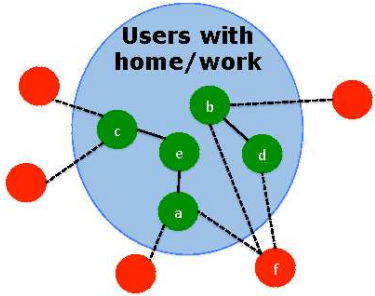}
\caption{How social filtering works. Green nodes are the ones for which we have identifies their home and work location. Red nodes are their neighbors (w/o identified home and work). We only consider ride-sharing among the green nodes. In the case of one-hop filtering, node \textit{a} can share a ride only with \textit{e}. In the case of two-hop filtering \textit{a} can share a ride with \textit{e},\textit{c},\textit{b}, and \textit{d}.}
\label{fig:graph_paradigm}
\end{figure}

\begin{figure}[t!]
\centering
\subfigure[call friends in Madrid]{\includegraphics[scale=0.195]{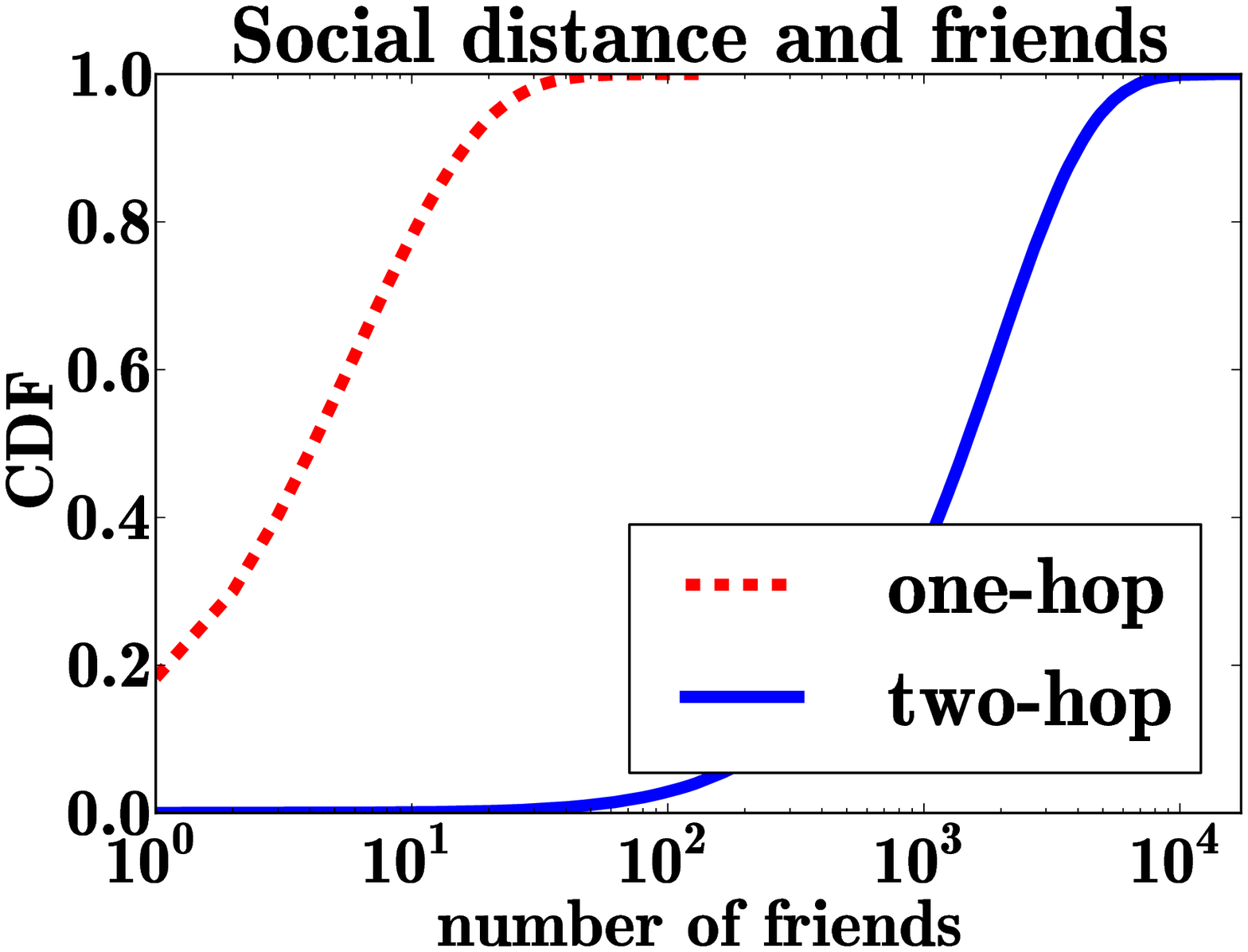}  \label{fig:useful-friends-madrid}}
\subfigure[declared friends in NY]{\includegraphics[scale=0.195]{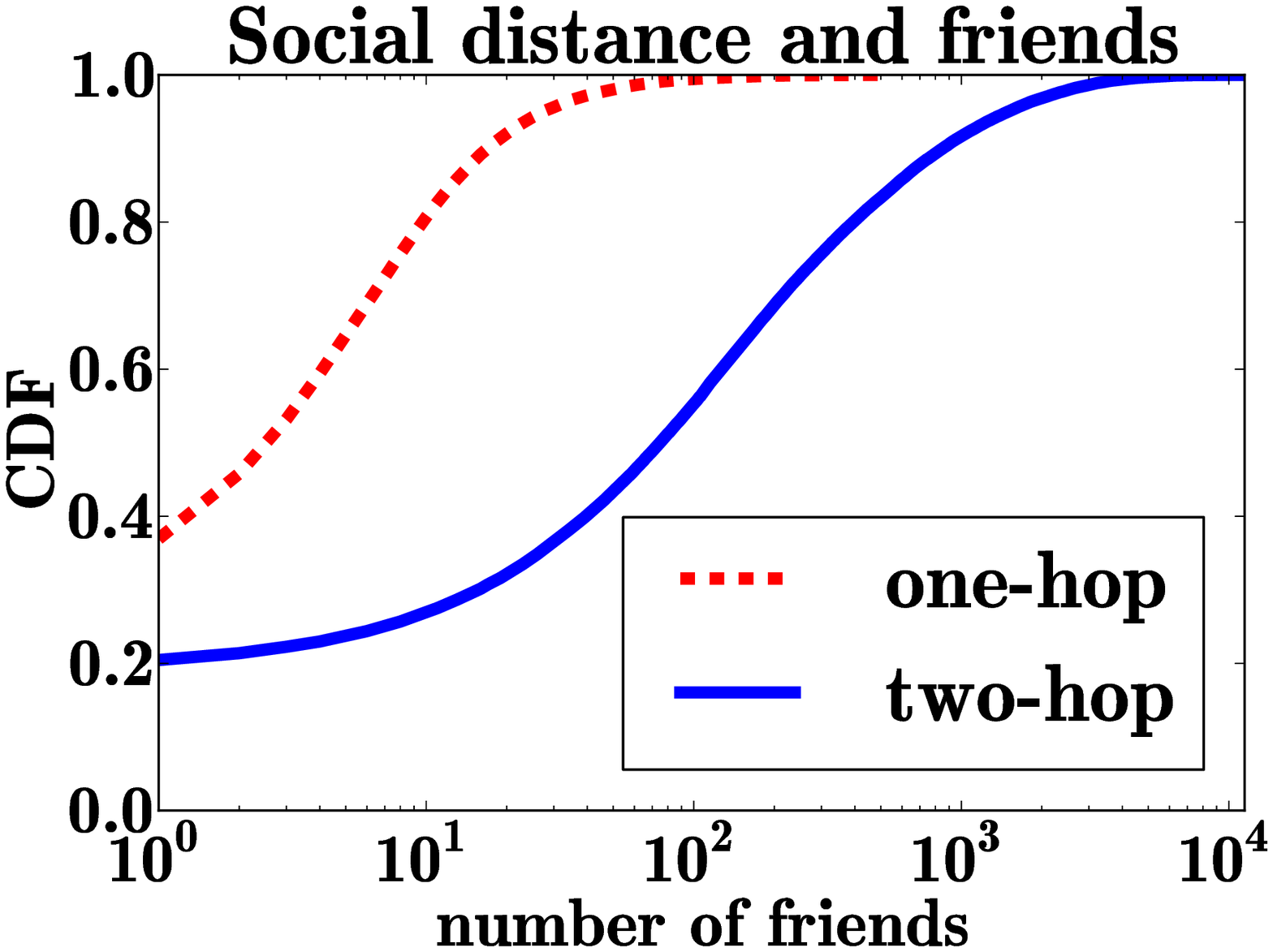}  \label{fig:useful-friends-twitter}}
\caption{Number of friends for the users with home and work address. X-axis is in log scale.}
\label{fig:useful-friends}
\end{figure}

In this section, we present how social filtering affects the potential of ride-sharing. Instead of assuming that anybody is willing to share a ride with anybody else, we introduce ``social constraints'' in selecting ride-sharing partners. The social constraints are represented by graphs, e.g. as shown in Fig. 15:  the nodes correspond to users, and the edges correspond to social ties between them. A user considers sharing a ride with a one-hop neighbor (i.e., somebody he knows directly, a ''friend'') or with a two-hop neighbor (a friend of a friend).

Given that we have two different types of data sets -- CDR and geo-tagged tweets -- we need to use two different definitions of edges. In the case of CDR data \cite{Dasgupta2008} \cite{Onnela2007}, choosing a ``threshold" condition for an edge between two users involves certain a trade-off between the strength of the tie and the number of edges. When choosing a threshold one needs to take into account the needs of the application \cite{Choudhury2010}. In this study, we create an edge in the social graph between two users when there is at least one call between them. We experimented with various definitions, and we found that  -- due to the small number users with inferred home/work locations -- higher thresholds would result in extremely sparse, thus useless,\footnote{Using a reciprocal call, as a threshold, would result in a graph with 2.4M nodes, and 3.7M edges. In that case, 92\% of the users had zero one-hop neighbors  with whom they could share a ride. As a result, the ride-sharing potential was 2\% (5.1\% with extrapolation) for \fullroute with 2-hop social filter, and $\delta = 1.0$ km (dist. constr.), $\tau$=10, $\sigma$ = 30 (time constr.)}  graphs.

In the case of Twitter, we crawl the friends and the followers of the users for whom we have home/work locations, and we create an edge in the social graphs iff there is bidirectional edge on Twitter. See Table~\ref{tab:our-graphs} for graph details. Moreover, to be sure that the friend nodes in our Twitter graph represent real people we considered only users who had at least one geo-tagged tweet. Finally, in both CDR and Twitter cases, we filtered out nodes with more than $1000$ friends, in order to exclude popular phone services, or celebrities, respectively. 

Fig.~\ref{fig:graph_paradigm} illustrates how filtering is done. The green nodes inside the circle, represent the users with identified home/work area, who are also the candidates for ride-sharing. The red nodes outside the circle represent users whose home/work areas remain unknown, but they have a social tie with one or more green nodes. Note that ride-sharing can occur only between green nodes.
 
Let's start with Madrid. As we can see from Table \ref{tab:soc-constr-data} the potential of ride-sharing is quite low when users are willing to share a ride only with their one-hop friends.
This is expected, since the graph shows only a small portion of a user's friends, and the users for whom we have home/work addresses are only a small subset of  all users. From Fig.~\ref{fig:useful-friends-madrid}, we can see that 80\% of the nodes in the call graph have no more than 10 one-hop friends, whose home/work addresses have been identified. But, if users are willing to share rides with friends of friends, then from Table~\ref{tab:soc-constr-data} we can see that, even with a sparse social graph, there can be considerable gain from \fullroute. This can be explained from Fig. ~\ref{fig:useful-friends-madrid}, in which we can see the much higher  number of two-hop than one-hop friends. In all data sets, there is a considerable improvement; e.g., in  Madrid, ride-sharing has a potential of 19\% (or 31\% with extrapolation to the entire population of Madrid). 

\begin{figure}[t!]
\centering
\subfigure[ Call graph ]{\includegraphics[scale=0.20]{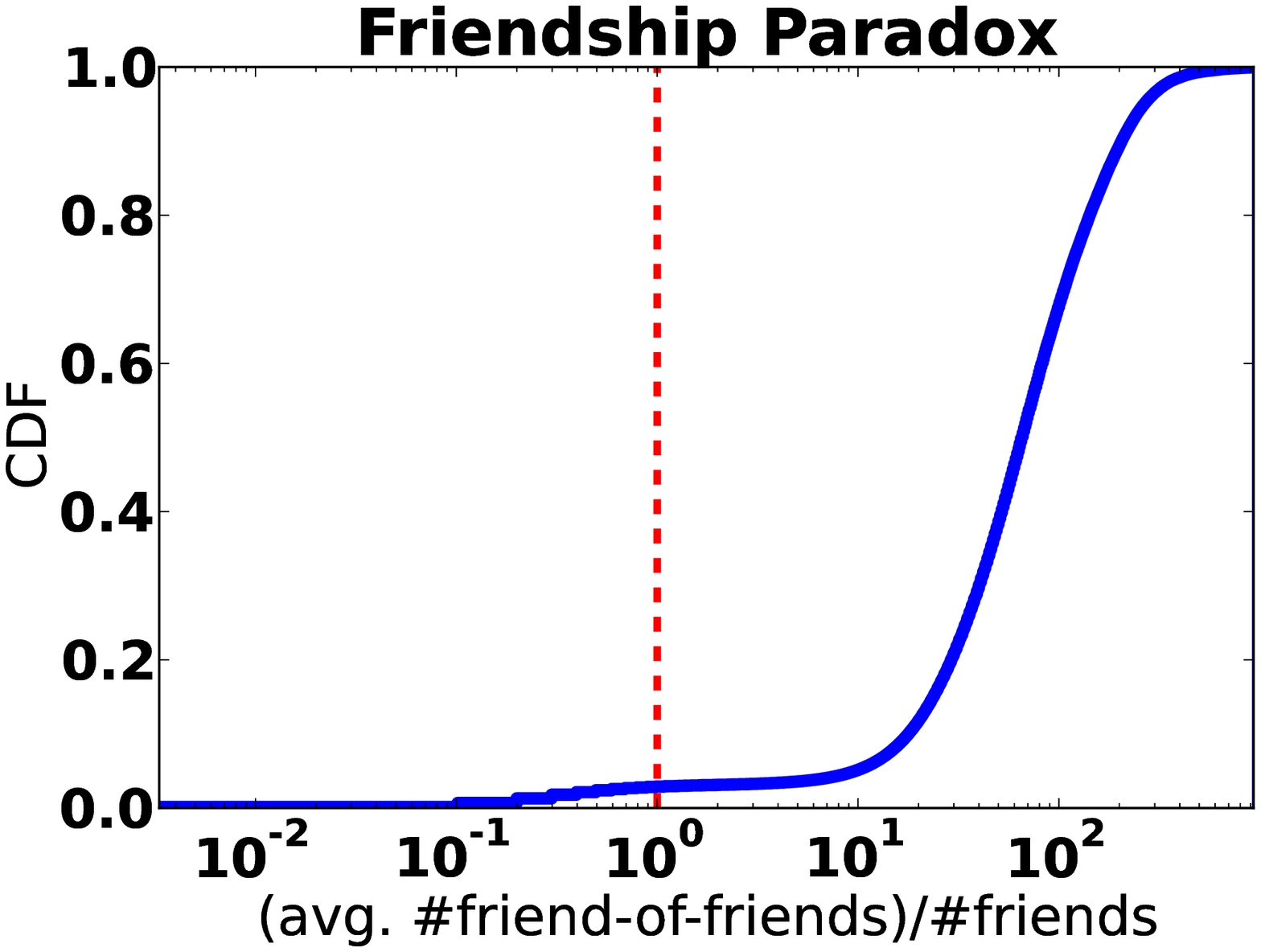}}
 \subfigure[Twitter graph]{\includegraphics[scale=0.20]{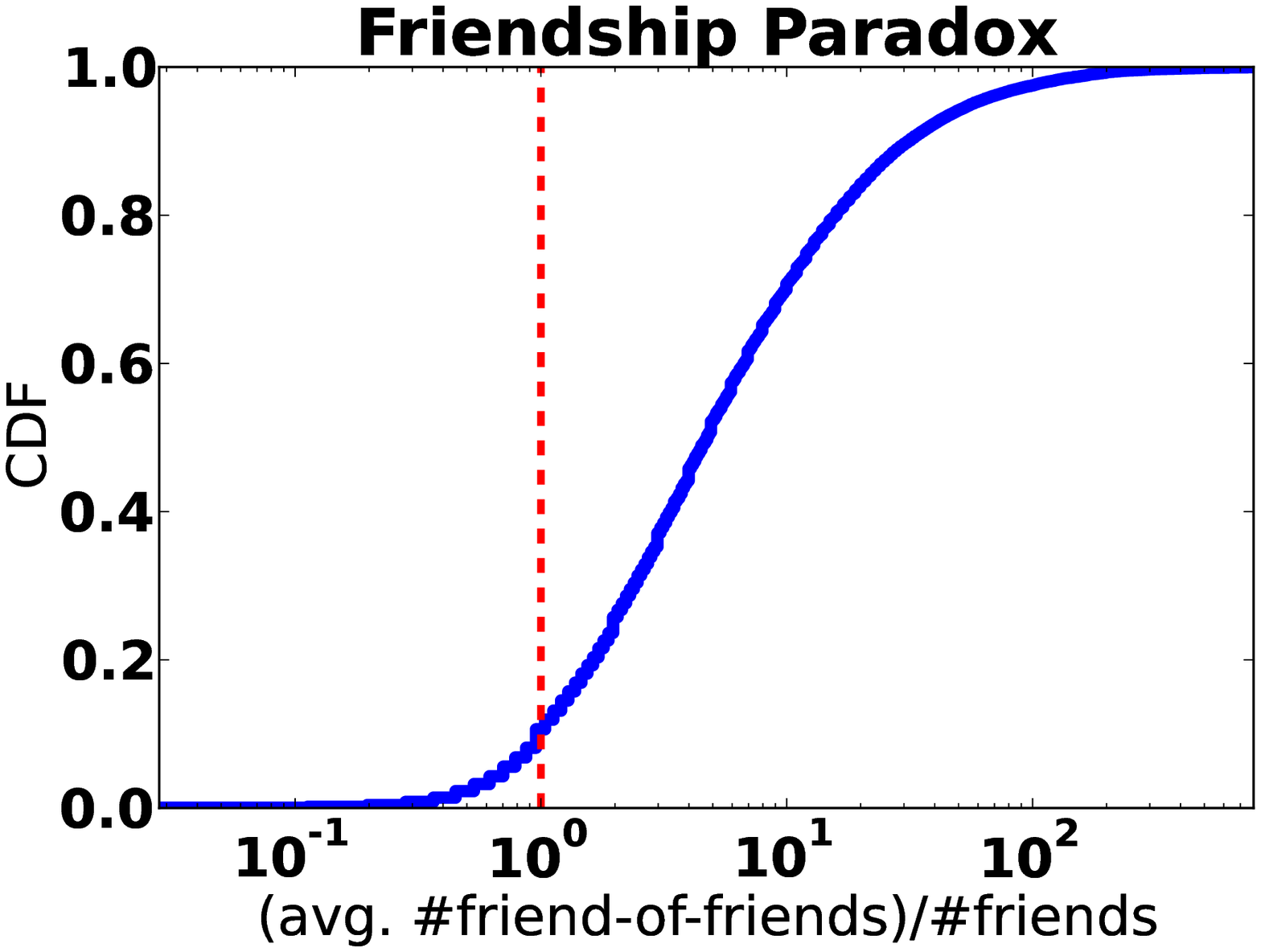}}
\caption{The CDF of the ratio between average \#friends-of-friends per \#friends. Friendship paradox holds when this ratio is greater than one (almost 90\% of the users both figures)}
\label{fig:friendship_paradox}
\end{figure}

In general, it can be observed that the number of nodes and edges in the social graph is crucial for any ride sharing application that wants to exploit social filtering. Moreover, the difference between the large increase in the ride-sharing potential when using friend-of-friends can be attributed to the friendship paradox ( ``on average your friends have more friends that you do'',  \cite{hodas13, feld1991} that also holds in our datasets as illustrated in Fig.~\ref{fig:friendship_paradox}.

%


\section{A tale of four cities}\label{sec:allcities}

\begin{table}[t!]
\caption {Madrid vs. Barcelona}
\small
\begin{tabular}{lcc}
\hline 
scenarios &\hwonly ratio. & \fullroute ratio. \\ 
 & (\%) & (\%) \\ \hline
$\tau$=10, $\sigma$ = 30 & 3.3 & 1.8 \\
Social constr. & 68 & 14  \\ \hline
\end{tabular}
\label{tab:madrid-vs-barca}
\\ This table shows the difference in ride-sharing potential between Barcelona and Madrid, for both \hwonly and 
\fullroute, in two different scenarios : (1) $\delta = 1.0$ km, $\tau$=10, and $\sigma$= 30, and (2) $\delta = 1.0$ km, $\tau$=10, $\sigma$= 30, and two-hop friends. 
 The ration is computed by : $((BCN - Madrid)/Madrid) * 100$
\end{table}

\begin{table}[t!]
\caption {NY vs. LA}
\small
\begin{tabular}{lcc}
\hline 
scenarios &\hwonly ratio. & \fullroute ratio. \\ 
 & (\%) & (\%) \\ \hline
$\tau$=10, $\sigma$ = 30 & -33 & -9 \\
Social constr. & -50 & -46 \\ \hline
\end{tabular}
\label{tab:ny-vs-la}
\\ This table shows the difference in ride-sharing potential between New York and Los Angeles, for both \hwonly and 
\fullroute, in two different scenarios : (1) $\delta = 1.0$ km, $\tau$=10, and $\sigma$=30, and (2) $\delta = 1.0$ km, $\tau$=10, $\sigma$= 30, and two-hop friends. The ration is computed by : $((LA - NY)/NY) * 100$
\end{table}

In this section, we compare the potential of ride-sharing in the four metropolitan areas (Madrid, Barcelona, New York, and Los Angeles) captured in our datasets. We discuss how the density of each city affects the potential or ride-sharing. 

\begin{figure}[t!]
\centering
\subfigure[Barcelona vs. Madrid, 2-hop friend distr.]{\includegraphics[scale=0.19]{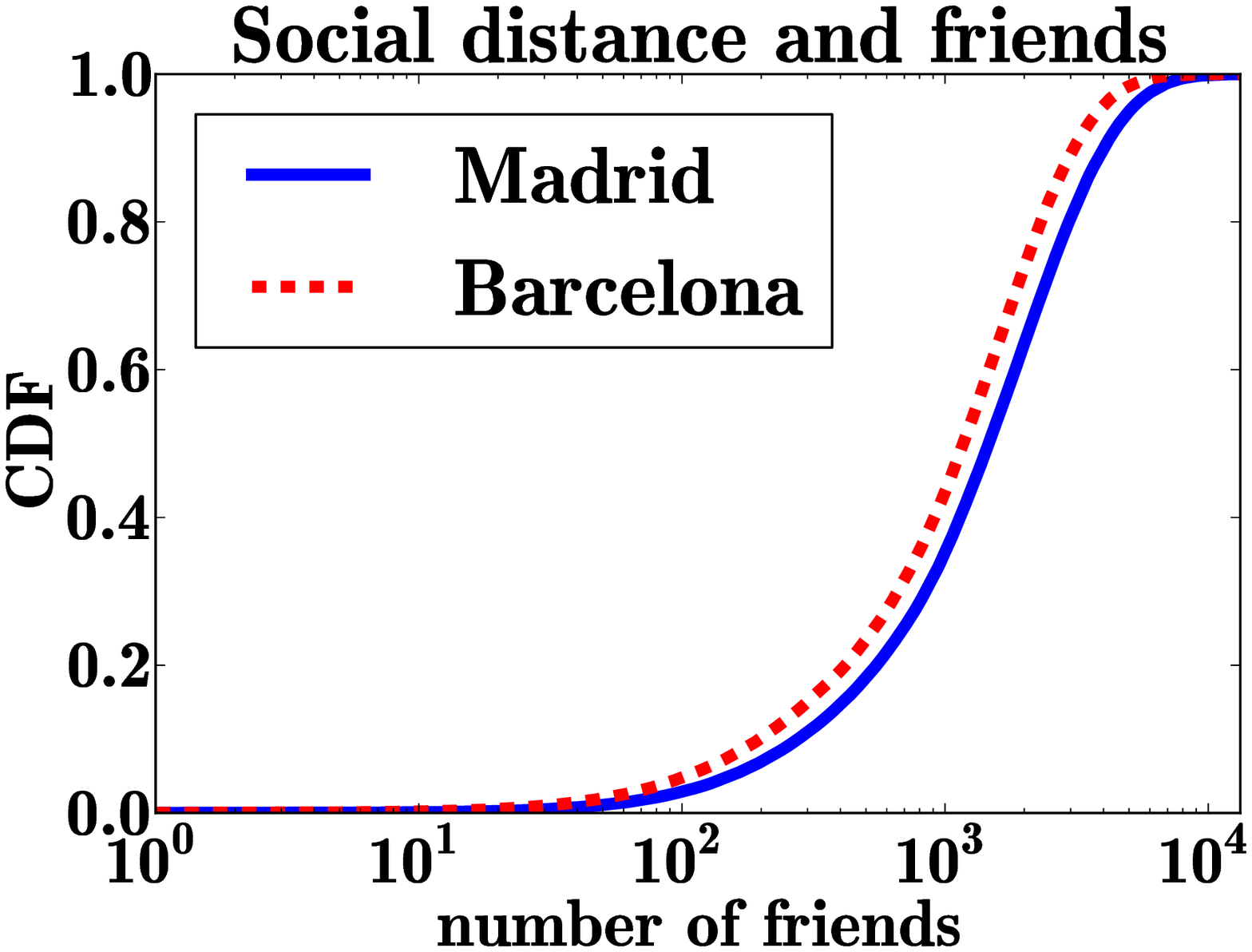}  \label{fig:madrid-vs-barca}}
\subfigure[NY vs. LA, 2-hop friend distr.]{\includegraphics[scale=0.19]{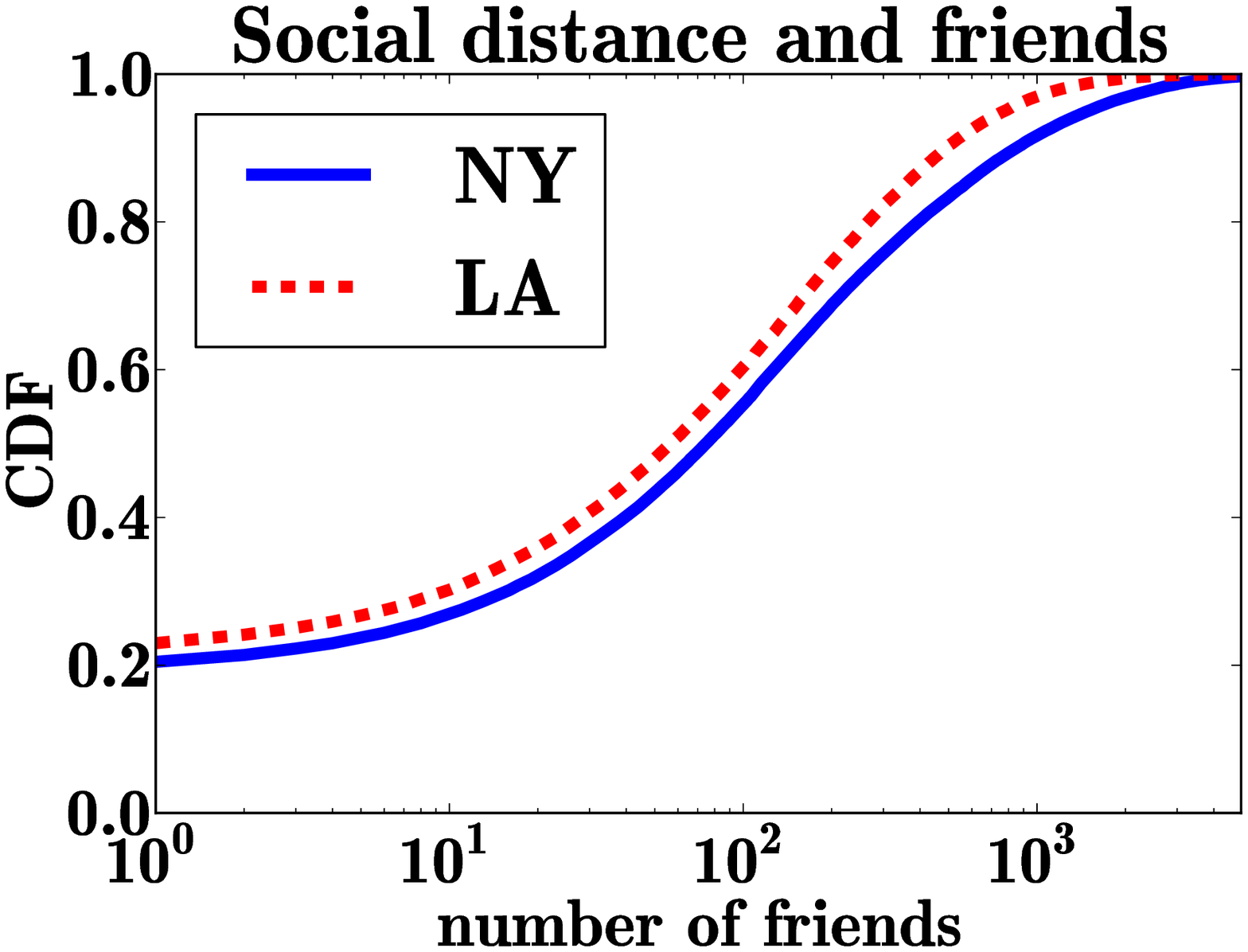}  \label{fig:ny-vs-la}}
\caption{Comparing social filtering between Madrid and Barcelona, and between NY and LA.}
\label{fig:city-social-compare}
\end{figure}

We start by comparing Madrid and Barcelona.  The first row of Table~\ref{tab:madrid-vs-barca} shows that, for spatio-temporal constraints only, the potential or ride-sharing in the two cities is very similar, with the potential of \fullroute being slightly higher in Barcelona. In the second row, we show that when also considering social constraints, the relative difference in ride-sharing benefit between the two cities becomes becomes much higher: the potential of \hwonly in Barcelona is 68\% higher, and the potential of \fullroute in Barcelona is 14\% higher.  This difference cannot be explained by the social graph, since, as we can see from Fig.~\ref{fig:madrid-vs-barca}, the users in both cities have almost the same number of friends. We attribute the better potential in Barcelona to its higher population density: Madrid has a density of $5,390/km^2$, while Barcelona has a density of $15,926/km^2$. 

The same observation holds in the comparison between the two US cities. The potential or ride-sharing in New York is higher that the potential of ride-sharing in Los Angeles-- see Table ~\ref{tab:ny-vs-la}. The difference gets even higher when time or social constraints are included -- see Table~\ref{tab:ny-vs-la}. Again, the difference in the potential of ride-sharing can be explained by the densities of the two cities: Los Angeles has a density of $3,124/km^2$, and New York has a density of $10,429/km^2$.

We obtained the mobility data of Madrid and Barcelona from CDRs, and we obtained the mobility data of New York and Los Angeles from geo-tagged Tweets, therefore a direct comparison between European and US cities may lead to incorrect conclusions. However, both comparisons (Madrid \vs Barcelona and New York \vs Los Angeles) show that ride-sharing is more beneficial in cities with higher density, especially when time and social constraints are considered.

\section{Summary and Conclusion}

We used mobile and social data to demonstrate that there is significant overlap in people's commute in a city that indicates 
a high potential benefits from ride-sharing systems. This is clearly an upper bound to any practical ride-sharing system, but the positive result motivates the deployment of such systems and policies. Our results indicate that en-route ride-sharing with up to two-hop social contacts offers a good trade-off between technological feasibility, people's security concerns, and a substantial impact on traffic reduction. A more detailed {\em summary of our findings} is as follows.


We started by considering  \hwonly in which rides can be shared only with neighbors in both home and work. Even with a modest radius of 1.0 km 
we observed a great potential reduction of cars. In the case of Madrid, this reduction is  59\%, based on our location data set that captures close to 8\% percent of the total population. Our estimation of the ride-sharing potential projected to the total commuting population of the city is significantly higher (see Sect.~\ref{subsec:enrouteresults}).
The distribution of home and work locations, which is far from the uniform, is crucial to the success of ride-sharing: if Madrid had a uniform home and work distribution then the reduction would be 13\% assuming only spatial constraints, and 0.2\% assuming time constraints too. This is in agreement with \cite{tsao99} and shows that ride-sharing has negligible benefit in a city with uniform home/work distribution. 

Adding time constraints, the effectiveness of ride-sharing becomes proportional to the driver/passenger waiting time for a pick-up, and inversely proportional to the standard deviation of the distribution of departure times. With a standard deviation of 30 min, a wait time up to 10 min and a $\delta$ of 1km there is a 24\% reduction of cars in Madrid. 

\fullroute, i.e., allowing passenger to be picked up along the way, yields a great boost in ride-sharing potential with or without time constraints. In the case of Madrid, \fullroute increases the savings from 24\% to 53\%. 

The previous results assumed that passengers and drivers can be matched based only on distance and time of commute. In reality, people are often hesitant to ride with strangers, which significantly ride-sharing opportunities. The CDR and Twitter data do provide information about whether users know each other, as indicated  by calls in the CDR data set, or by a declared friendship in Twitter (Section 6). First, we consider ride-sharing only with one-hop friends. Then \fullroute in the city of Madrid using CDR and Twitter  friendship provides only a tiny traffic reduction of 1.1\% and 1.2\% respectively. This dramatic decrease is attributed to the low density of the social graphs and to the fact that only a small portion of the graphs' nodes have known home/work addresses -- each user has the opportunity to share a ride only with a small portion of her neighbors. However, if we relax  the social constraints and permit ride-sharing with ``friends-of-friends'', the ride-sharing potential increases significantly, especially in \fullroute. The corresponding numbers are 19\%  and 8.2\% for friendship based on CDRs and  Twitter data, respectively. Furthermore, if we project the potential of ride-sharing to the total commuting population of the city (which is much larger than the number of users with known home and work location in our data),  the benefit increases up to 31\% for call based filtering and 26\% for OSN based filtering.

Finally, we compared the four cities and observed some differences in Section 7. The population density of a city can have a profound effect on its ride-sharing potential, especially when strict social filtering is applied. For example,  Barcelona is denser and has a 14\% higher ride-sharing potential than Madrid; Los Angeles, on the other hand, has 46\% lower ride-sharing potential than New York.


Directions for future work include designing real-time matching algorithms (motivated by the offline analysis in this paper) 
 and implementing a prototype ride-sharing system. The methodology developed in this paper can potentially be used on other cities and different data sets to assess the inherent potential of ride-sharing and guide related deployment and policies.

\vspace{7mm}

\bibliographystyle{ieeetr}
\bibliography{wwwbp}


\end{document}